\documentclass{article}

\usepackage{amsfonts}
\usepackage{amsmath}
\usepackage{xypic}

\title{Algebraic properties of three-four dimensional anisotropic oscillator potentials through Hamiltonian chains.}
\author{Y. Tanoudis \\ Mathematics Department\\ Aristotle University of Thessaloniki\\ 54124 Greece \\ { E-mail: tanoudis@math.auth.gr}}

\begin{document}

\maketitle

\begin{abstract}
In this work the notion of \emph{Hamiltonian chain} is presented as applied to anisotropic oscillator potentials especially defined on three and four dimensional Euclidean spaces. A Hamiltonian chain is a sequence of superintegrable Hamiltonians which, in addition, constitute integrals of motion of a new superintegrable system. Along with the Poisson algebras the quantum counterparts of the chain is given as well as the eigenvalues of each member of the chain. The method can be extended to cover  $n$ dimensional cases.

\end{abstract}

\section{Introduction}

In classical and quantum mechanics an integrable system defined on a $N$-dimensional manifold, is a system which has $N$ integrals of motion in evolution. A superintegrable (or maximally integrable) system is a system possessing the maximum number of constants of motion, i.e. $2 N -1.$

The two-dimensional  superintegrable systems, with  integrals of motion which are quadratic polynomials of momenta, are classified by using the quadratic associative algebra of the integrals \cite{DasYps06, DasTan07, DasTan08}. In addition the problem of superintegrable systems with integrals of order 3 and higher is a topic of major interest studied in \cite{GrWin, Grav,TemWinMiPog,MarWin07JMP48, MarWin08JPA41, Mar09JMP50a, Mar09JMP50b,Mar09JMP50c,TrTurWin,Mar10JPA43,PosWin}. Significant work have been made, also, in three dimensional potentials \cite{KalKrMi07, KalKrMi07JPA, VerEvans08, TanDas08, DaskTan10, TanDas11}.  Several cases of three dimensional superintegrable systems with quadratic integrals of motion are described and analyzed by Kalnins, Kress and Miller \cite{KalKrMi07,KalKrMi07JPA}. Specifically, Kalnins, Kress and Miller studied a special case of superintegrable systems in which the potentials depend of four free parameters, these systems are  referred  as  \emph{non degenerate potentials}. Towards the $n$-dimensional systems Ballesteros and  Herranz in \cite{BalHer} studied specific families of superintegrable systems with quadratic integrals . Furtehrmore  Hoque, Marquette and  Zhang have introduced some families of $n$-dimensional potentials in \cite{HoMarZha15a, HoMarZha15b, HoMarZha15c}.

In \cite{KalKrMi10} Kalnins, Kress, and Miller have described a method for determining a complete set of integrals for a classical Hamiltonian that separates in orthogonal subgroup coordinates. This work was inspired by the work of Tremblay, Turbiner and Winternitz \cite{TeTurWin} concerning an infinite family of exactly solvable quantum mechanical systems denoted by (TTW). The proposition that every member of this family of systems is superintegrable, while not proven, is supported by the fact that their classical counterparts have been shown to be superintegrable \cite{KMiPog}. Kalnins, Kress and Miller using methods described in \cite{KMiPog, KKMPog} have demonstrated that some natural $n$ dimensional generalizations of the (TTW) class of classical systems are superintegrable, that is, possess $2 n -1$ constants that are polynomial in the momenta.
All superintegrable systems previously known, exist on conformally flat spaces. However, the underlying metrics of the new systems presented in \cite{KalKrMi10} are not necessarily conformally flat. The work of Kalnins, Kress and Miller in \cite{KalKrMi10} begins considering the next $n$ functions
\[
L_i = p_i^2 + V_i(q_i) + f_i(q_i) L_{i+1},\quad i=1,\ldots,n-1
\]
\[
L_n = p_n^2 + V_n(q_n)
\]
on a $2 n$ dimensional phase space with position coordinates $q_i$ and conjugate momenta $p_i$, $i=1,\ldots,n$. Each $L_i$ is a function of $q_1,\ldots,q_n$,$p_1,\ldots,p_n$ and the Hamiltonian $H=L_1$.
Furthermore the set $\{L_1,\ldots,L_n\}$ is in involution that is $\{L_i,L_j\}_P=0$ for all $i,j=1,\ldots,n$, where $\{,\}_P$ the usual Poisson bracket,
\[
\{L_i,L_j\}_P = \sum_{i=1}^{n} \frac{\partial L_i}{\partial q_k}\frac{\partial L_j}{\partial p_k} - \frac{\partial L_i}{\partial p_k}\frac{\partial L_j}{\partial q_k}
\]
We can find $n-1$ additional functionally independent functions $L'$ commuting with the Hamiltonian $H$ given by the following relations
\[
L_i' = N_i -M_i, \quad i=1,\ldots,n-1
\]
where
\[
M_i = \int \frac{f_i(q_i)}{2 \sqrt{L_i-V_i(q_i)-f_i(q_i)L_{i+1}}} \;dq_i
\]
\[
N_i = \int \frac{dq_i}{2 \sqrt{L_{i+1}-V_{i+1}(q_{i+1})-f_{i+1}(q_{i+1})L_{i+2}}}
\]
This method is also used in \cite{KMi12} considering an infinite class of classical extended Kepler-Coulomb 3- and 4-parameter systems.
The algebraic structure of these families of superintegrable systems is still an open problem both in specific systems as well as in general forms.

Tempesta, Turbiner and Winternitz in \cite{TeTurWin}  and Kalnins, Kress and Miller in \cite{KalKrMi10} have introduced a chain of multidimensional  superintegrable Hamiltonians generated by the
recurrence formula
\[
\begin{array}{l}
H_1=p_1^2 + V_1(q_1)\\
H_i=p_i^2 + V_i(q_i)+\frac{1}{\phi_i(q_i)} H_{i-1}
\end{array}
\]

In present work the study of both triple Poisson algebras and triple Lie-Poisson algebras of the last mentioned Hamiltonian chain are presented.  Especially the case of anisotropic oscillator potential is used in order to produce a classification of this chain of superintegrable Hamiltonians. The three dimensional case of generalized anisotropic oscillator potential is written as follows
\[
H= \frac{1}{2} ( p_x^2+ p_y^2 +p_z^2)+ k (a_1^2 x^2+ a_2^2 y^2 + a_3^2 z^2) + \frac{k_1}{x^2} +  \frac{k_2}{y^2}+  \frac{k_3}{z^2}
\]

In addition to the quantum triple Lie-Poisson algebras the computation of the eigenvalues for the involved integrals is given using the deformed oscillator technique \cite{Das01} as applied to higher dimension superintegrable systems\cite{TanDas11b}.

\section{The `c' and `q' oscillator systems in classical mechanics. }

Let us consider the next two-dimensional, anisotropic oscillator, Hamiltonian function $\mathcal{H}_1$ with position coordinates $x_i$ and conjugate momenta $p_{x_i}$
\begin{equation}\label{eq:CubHamiltonian}
\mathcal{H}_1 = \frac{1}{2} \left( p_{x_1}^2 +p_{x_2}^2 \right ) + \kappa \left( x_1^2 + 4 x_2^2 \right ) + \frac{\kappa_1}{x_1^2} + \frac{\kappa_2}{x_2^2} + r
\end{equation}
where $\kappa, \kappa_i, r$ arbitrary constants.

The Hamiltonian (\ref{eq:CubHamiltonian}) admits two functionally independent integrals of motion denoted by $A_1, B_1$ which given as follows
\begin{equation}\label{eq:CubA1}
A_1 = \frac{1}{2} p_{x_1}^2 +  \kappa x_1^2 + \frac{\kappa_1}{x_1^2}+ r
\end{equation}
\begin{equation}\label{eq:CubB1}
\begin{split}
B_1 = p_{x_1}^2 J_3^2  + & \left ( 2 \kappa_2 \frac{x_1^2}{x_2^2} + 4 \kappa_1 \frac{x_2^2}{x_1^2}  - 4 \kappa x_1^2 x_2^2 \right )p_{x_1}^2+ \\& \hspace{7em} + \left ( 4 \kappa x_1^3 x_2  - 4 \kappa_1 \frac{x_2}{x_1}  \right ) p_{x_1} p_{x_2}   + \frac{4 x_2^2 (\kappa x_1^4 - \kappa_1)^2}{x_1^4}
\end{split}
\end{equation}
where $J_3= x_1 p_{x_2}  - x_2 p_{x_1}.$

The set of integrals $(\mathcal{H}_1,A_1,B_1)$ constitutes a two-dimensional superintegrable system.  This system possesses one fourth order integral $B_1$  among the others $\mathcal{H}_1, A_1$  which are second order polynomial functions in momenta. Moreover, the above system forms a cubic triple Poisson algebra the general form of which is given as follows
\begin{equation}
\label{eq:cubicclassalgebra}
\begin{split}
  \{A_1,B_1\}_P = C_1 \hspace{11em}\\
\{A_1,\{A_1,B_1\}_P\}_P = a A_1^3 + \alpha A_1^2 + \beta \{A_1,B_1\} + \gamma A_1 + \delta B_1 + \epsilon \hspace{1em}\\
\{B_1,\{A_1,B_1\}_P\}_P = \mu A_1^3 + \nu A_1^2 - \beta B_1^2 - a \{A_1,A_1,B_1\} - \alpha \{A_1,B_1\} + \\  + \xi A_1 -\gamma B_1 + \zeta
\end{split}
\end{equation}
where $\{A,A,B\}= A^2 B + A B A + B A^2$ and  $\{A, B\} =  A B + B A$.  The general form of the algebra coefficients is
\[
\alpha = \alpha_1 \mathcal{H}_1 + \alpha_0, \; \gamma = \gamma_2 \mathcal{H}_1^2 + \gamma_1 \mathcal{H}_1 + \gamma_0, \;  \delta = \delta_1 \mathcal{H}_1 + \delta_0
\]
\[
\epsilon = \epsilon_3 \mathcal{H}_1^3 + \epsilon_2 \mathcal{H}_1^2 +\epsilon_1 \mathcal{H}_1 + \epsilon_0, \; \mu = \mu_1 \mathcal{H}_1 + \mu_0, \; \nu = \nu_2 \mathcal{H}_1^2 + \nu_1 \mathcal{H}_1 + \nu_0
\]
\begin{equation}\label{eq:CubClasStrucConstGener}
\xi = \xi_3 \mathcal{H}_1^3 + \xi_2 \mathcal{H}_1^2 + \xi_1 \mathcal{H}_1 +\xi_0, \; \zeta = \zeta_4 \mathcal{H}_1^4 + \zeta_3 \mathcal{H}_1^3 + \zeta_2 \mathcal{H}_1^2 + \zeta_1 \mathcal{H}_1 + \zeta_0
\end{equation}
where $a, \beta, \alpha_i, \gamma_i, \delta_i, \epsilon_i, \mu_i, \nu_i, \xi_i, \zeta_i $ arbitrary constants and $\mathcal{H}_1$ the Hamiltonian function of the system. Furthermore, the algebra (\ref{eq:cubicclassalgebra}) possesses a Casimir $K_1$ which is a polynomial function of the integrals $A_1, B_1$ in order the following relations to be true
\[
\{K_1,\mathcal{H}_1\}_P=\{K_1, A_1\}_P = \{K_1, B_1\}_P = 0
\]
The general formula for the Casimir $K_1=K_1(A_1,B_1)$, when the algebra coefficients (\ref{eq:CubClasStrucConstGener}) are known,  is given by the following relation
\begin{equation}\label{eq:cubicCasimir}
\begin{split}
 K_1=  C_1^2   -\frac{a}{2}\{A_1,A_1,A_1,B_1\} - \frac{2 \alpha}{3} \{A_1,A_1,B_1\} -\frac{2 \beta}{3} \{A_1,B_1,B_1\} -  \\   -  \gamma \{A_1,B_1\} +   \frac{2 \nu}{3}  A_1^3 -\delta  B_1^2 + \frac{\mu}{2} A_1^4  -2 \epsilon   B_1 + \xi  A_1^2   + 2 \zeta A_1 =\\  = k_0 + k_1 \mathcal{H}_1 + k_2 \mathcal{H}_1^2 + k_3 \mathcal{H}_1^3 + k_4 \mathcal{H}_1^4 + k_5 \mathcal{H}_1^5
\end{split}
\end{equation}
where $\{A_1,A_1,A_1,B_1\} = A_1^3 B_1 + A_1^2 B_1 A_1 + A_1 B_1 A_1^2 + B_1 A_1^3.$
Conversely the algebra (\ref{eq:cubicclassalgebra}) is generated by the Casimir (\ref{eq:cubicCasimir}) as follows
\begin{equation}
\begin{split}
&\{A_1,B_1\}_P = \frac{1}{2}\frac{\partial K_1}{\partial C_1}, \\
 \{B_1,\{A_1,B_1\}_P\}_P = &\frac{1}{2} \frac{\partial K_1}{\partial A_1}, \{\{A_1,B_1\}_P, A_1\}_P =  \frac{1}{2}\frac{\partial K_1}{\partial B_1}
\end{split}
\end{equation}
The expression   (\ref{eq:cubicCasimir}), also, can be written
\[
K_1 = C_1^2 - 2 h^{c}
\]
where $h^{c}$ the generating function.

The system  (\ref{eq:CubHamiltonian})-(\ref{eq:CubB1}) will be refereed, for simplicity reasons, as a \emph{c-system}. The corresponding generating function is
\begin{equation}\label{eq:generatcubgener}
\begin{split}
 h^{c} =  \frac{16}{3} (H_1-r) \{ (A_1-r),(A_1-r), B_1 \} + 32 \kappa \kappa_1 \{(A_1-r), B_1\}   + \\ - 4 \{(A_1-r),(A_1-r),(A_1-r), B_1 \} - 64 \kappa_2 (A_1-r)^4 +\\ + 512 \kappa \kappa_1 \kappa_2 (A_1-r)^2  - 16 \kappa B_1^2 - 64 \kappa \kappa_1 (H_1-r) B_1 - \\ - 512 \kappa \kappa_1 \kappa_2 r^2 + 64 \kappa_2 r^4
\end{split}
\end{equation}
In case of the \emph{c-system} the coefficients $k_i$ are
\[
k_1=k_2=k_3 = k_4 = k_5 = 0, \quad k_0 = -128 \kappa_2 (r^2 - 4 \kappa \kappa_1)^2
\]

Additionally, we consider the following  two-dimensional, isotropic oscillator, Hamiltonian function $\mathcal{H}_2$ with position coordinates $q_i$ and conjugate momenta $p_{q_i}$.

\begin{equation}\label{eq:QuadHamiltonian}
\mathcal{H}_2 = \frac{1}{2} \left( p_{q_1}^2 +p_{q_2}^2 \right ) + \lambda \left( q_1^2 +q_2^2 \right ) + \frac{\lambda_1}{q_1^2} + \frac{\lambda_2}{q_2^2} + s
\end{equation}
where $\lambda, \lambda_i, s$ arbitrary constants.
The last Hamiltonian admits two functionally independent integrals of motion denoted by $A_2, B_2$ given as follows
\begin{equation}\label{eq:QuadA2B2}
A_2 = \frac{1}{2} p_{q_2}^2 + \lambda q_2^2 + \frac{\lambda_2}{q_2^2}+ s, \quad B_2 = J_3^2 + \frac{2 \lambda_1 q_2^2}{q_1^2} + \frac{2 \lambda_2 q_1^2}{q_2^2}
\end{equation}
where $J_3=q_1 p_{q_2} - q_2 p_{q_1}.$

This system it is a well known superintegrable system admitting a triple qua\-dra\-tic Poisson algebra given by the following relations
\begin{equation}
\label{eq:quadraclassalgebra}
\begin{split}
& \hspace{-2em} \{ A_2, B_2\}_P=C_2,  \\
 \{ A_2, \{A_2, B_2\}\}_P = \alpha A_2^2 + \beta B_2^2 &
+ \gamma  \{A_2, B_2\}  + \delta A_2 + \epsilon B_2 + \zeta,\\
\{ B_2, \{A_2, B_2\}\}_P = a A_2^2 - \gamma B_2^2  & - \alpha \{ A_2, B_2\} + d A_2
-\delta B_2 + z.
\end{split}
\end{equation}
The algebra coefficients are given as follows
\begin{equation}
\begin{split}
& \delta=\delta_{1} \mathcal{H}_2 +\delta_{0}, \quad
 \epsilon=\epsilon_{1} \mathcal{H}_2 +\epsilon_{0}, \\
 \zeta=\zeta_{2} \mathcal{H}_2^2+\zeta_{1} \mathcal{H}_2 & +\zeta_{0}, \quad
 d=d_{1} \mathcal{H}_2+d_{0}, \quad
 z=z_{2} \mathcal{H}_2^2+z_{1} \mathcal{H}_2+z_{0,}
\end{split}
\end{equation}
where $a, \alpha, \beta, \gamma, \delta_i, \epsilon_i, d_i, \zeta_i, z_i$ constants.
The above structure (\ref{eq:quadraclassalgebra}) possesses a Casimir $K_2(A_2,B_2)$  which in general is given by the following relation
\begin{equation}\label{eq:QuadraticCasimir}
\begin{split}
K_2 = C_2^2 + \frac{2 a}{3} A_2^3 - \frac{2 \alpha}{3} \{A_2,A_2,B_2\} - \frac{2 \gamma}{3} \{A_2,B_2,B_2\}  -\delta \{A_2,B_2\} + \\ +   d  A_2^2   - \epsilon B_2^2 +  2 z A_2  - 2 \zeta  B_2  = k_0 + k_1 \mathcal{H}_2 + k_2 \mathcal{H}_2^2 + k_3 \mathcal{H}_2^3
\end{split}
\end{equation}
Furthermore the algebra (\ref{eq:quadraclassalgebra})  is generated by the Casimir (\ref{eq:QuadraticCasimir}) as follows
\begin{equation}
\begin{split}
&\{A_2,B_2\}_P = \frac{1}{2}\frac{\partial K_2}{\partial C_2}, \\
\{B_2,\{A_2,B_2\}_P\}_P = & \frac{1}{2} \frac{\partial K_2}{\partial A_2}, \{\{A_2,B_2\}_P,A_2\}_P =  \frac{1}{2}\frac{\partial K_2}{\partial B_2}
\end{split}
\end{equation}

The expression   (\ref{eq:QuadraticCasimir})  can be written also,
\[
K_2 = C_2^2 - 2 h^{q}
\]

From now on we shall refereed to the system (\ref{eq:QuadHamiltonian})-(\ref{eq:QuadA2B2})  as a \emph{q-system}. The generating function is
\begin{equation}\label{eq:generatquadrgener}
\begin{split}
h^{q} = -\frac{8}{3} \{ (A_2-s),(A_2-s),B_2\} + 4 (\mathcal{H}_2-s) \{(A_2-s),B_2\} - 4 \lambda B_2^2 - \\ -16 (\lambda_1 + \lambda_2) (A_2-s)^2 + 32 \lambda_2 (\mathcal{H}_2-s)(A_2-s) + \\ +32 \lambda_2 (\mathcal{H}_2-s) s + 16 (\lambda_1+\lambda_2) s^2
\end{split}
\end{equation}
Moreover the coefficients $k_i$ are
\[
k_2= -32 \lambda_2,\; k_0= 128 \lambda \lambda_1 \lambda_2 -32 \lambda_1 s^2, \; k_1=k_3 = 0
\]
or
\[
K_2 = -32 \lambda_2 (\mathcal{H}_2-s)^2 -64 \lambda_2 s (\mathcal{H}_2-s) + 128 \lambda \lambda_1 \lambda_2 -32 (\lambda_1+\lambda_2) s^2
\]

Both systems can be presented by the following diagram
\begin{equation}\label{eq:diagram}
\xymatrix{ A_i \ar@{--}[r] & E_i \ar@{--}[r] & B_i}
 \end{equation}
 where the dashed line joining two integrals  means that the corresponding commutator is zero, the absence of any joining line between $A_i$ and $B_i$ means that the corresponding commutator is different to zero.

\section{The quantum form of the `q' system. }\label{sec:eigenqsys}

Any two-dimensional quantum superintegrable system with integrals quadratic polynomials in momenta is described by the Hamiltonian $H$ and two functionally independent integrals of motion $A$ and $B$. The integrals $A$ and $B$ commute with Hamiltonian $H$, but they don't commute between them
\[
\left[H,\, A\right]= 0, \quad \left[H,\, B\right]= 0  \quad \mbox{ and }\quad \left[A,\, B\right]\ne 0
\]
Let $\mathcal{A}=\mathbb{C}\left<A,B,H\right>$ be the unital algebra generated by the operators $A,\,B,\, H$, the generators of this algebra satisfy  ternary relations, which are quadratic extensions of the  enveloping algebra of a Lie triple system \cite{Jac51}.
\begin{equation}\label{eq:QLTS}
 \begin{array}{l}
\left[ A,\left[A,B\right] \right ]=\alpha A^2+\beta B^2+ \gamma \{ A,B \}+\delta A+ \epsilon B +\zeta \\
 \left [B,  \left[A,B\right] \right ] = a A^2-\gamma B^2-\alpha  \{ A, B\}  +d
A-\delta B+z
 \end{array}
 \end{equation}
where $\{A,B\}=A B + B A$. In a Lie triple system  \cite{Jac51} in the right hand side of the above equations  are only linear functions of the operators $A$ and $B$. In the case of superintegrable systems with quadratic integrals of motion there are also quadratic terms.  Some of the coefficients of the ternary quadratic algebra (\ref{eq:QLTS}) depend generally on the energy $H$.
\begin{equation}\label{eq:QLTScoef}
\begin{split}
& \delta=\delta_{1} H+\delta_{0}, \quad
 \epsilon=\epsilon_{1} H +\epsilon_{0}, \\
 \zeta=\zeta_{2} H^2+\zeta_{1} H & +\zeta_{0}, \quad
 d=d_{1} H+d_{0}, \quad
 z=z_{2} H^2+z_{1} H+z_{0,}
\end{split}
\end{equation}
We are interested to calculate the energy eigenvalues of the operator $H$,  therefore we search to calculate the values of the energy corresponding to finite dimensional representations of the algebra (\ref{eq:QLTS}).
There is a  Casimir $[K,A]=[K,B]=0$  and  $K=K(H)$
\begin{equation}\label{eq:QLTSCas}
\begin{split}
K=&\left[A,B\right]^2-\alpha \{ A^2,B\}-\gamma\{ A,B^2 \}+\left(\alpha
\gamma -\delta+\frac{a \beta}{3}\right)\{A,B\}-\frac{2 \beta}{3} B^3
\\&
+\left(\gamma^2-\epsilon-\dfrac{\alpha \beta}{3}\right) B^2+ \left(-\gamma
\delta+2\zeta-\frac{\beta  d}{3}\right)B   +\frac{2 a}{3} A^3+ \\+&\left(d+\frac{a
\gamma}{3}+\alpha^2\right)
A^2+\left(\frac{a \epsilon}{3}+\alpha \delta+2 z\right) A =\\
\\&=
h_0  + h_1 H + h_2 H^2 + h_3 H^3
\end{split}
\end{equation}

The quantum analog of the \emph{q-system} is given by the following expressions
\[
\mathcal{H}_2 = \frac{1}{2} \left( p_{q_1}^2 +p_{q_2}^2 \right ) + \lambda \left( q_1^2 +q_2^2 \right ) + \frac{\lambda_1}{q_1^2} + \frac{\lambda_2}{q_2^2} + s
\]
\begin{equation}\label{eq:quantumqsystem}
A_2 = \frac{1}{2} p_{q_2}^2 + \lambda q_2^2 + \frac{\lambda_2}{q_2^2}+ s
\end{equation}
\[
B_2 = J_3^2 + \frac{2 \lambda_1 q_2^2}{q_1^2} + \frac{2 \lambda_2 q_1^2}{q_2^2}
\]
where
\[
p_{q_1} = - i \hbar \partial_{q_1}, \; p_{q_2} = - i \hbar \partial_{q_2}, \; J_3 = - i \hbar (q_1 \partial_{q_2}-q_2 \partial_{q_1})
\]
In this case the full algebra is given as follows
\begin{equation}\label{eq:AABGenQuanQsyst}
[A_2,[A_2,B_2]] = 8 \hbar^2 (A_2-s)^2 - 8 \hbar^2 (\mathcal{H}_2-s) (A_2-s) + 8 \lambda \hbar^2 B_2 + 4 \hbar^4 \lambda
\end{equation}
\begin{equation}\label{eq:BABGenQuanQsyst}
\begin{split}
[B_2,[A_2,B_2]] = -16\bigl ( 2\hbar^2 (\lambda_1+\lambda_2)-\hbar^4 \bigr ) (A_2-s) - 8\hbar^2 \{(A_2-s),B_2\} + \\ +  8 \hbar^2 (\mathcal{H}_2-s) B_2 + 8 (  4 \hbar^2 \lambda_2 - \hbar^4 ) (\mathcal{H}_2-s)
\end{split}
\end{equation}
and the structure constants, based on the general form (\ref{eq:QLTS}), are
\[
\alpha = 8\hbar^2,\; \beta =0,\; \gamma = 0,\;  \delta = - 8\hbar^2 (\mathcal{H}_2-s) - 16 \hbar^2 s, \; \epsilon= 8 \hbar^2 \lambda,
\]
\[
 \zeta = 8 \hbar^2(\mathcal{H}_2-s) s + 8 \hbar^2 s^2 +4 \hbar^4 \lambda , \;  a=0,\;d = - 32 \hbar^2 (\lambda_1+\lambda_2) + 16 \hbar^4,
\]
\[
 z = 8 ( 4 \hbar^2 \lambda_2 -\hbar^4  )(\mathcal{H}_2 - s)  + 8  \bigl ( 4 \hbar^2(\lambda_1+\lambda_2)  - 2 \hbar^4 \bigr ) s
\]
Moreover, the Casimir invariant is given by the next formula
\begin{equation} \label{eq:qsystCasim}
\begin{split}
K = 4 (8 \hbar^2 \lambda_2 - 3 \hbar^4) (\mathcal{H}_2-s)^2 - 16 \Bigl ( 8 \hbar^2 \lambda \lambda_1 \lambda_2 - 3 \hbar^4 \lambda (\lambda_1+\lambda_2) +  \hbar^6 \lambda \Bigr ) + \\ + (64 \hbar^2 \lambda_2 - 80 \hbar^4) (\mathcal{H}_2-s) s +  \bigl(32 \hbar^2 (\lambda_1+\lambda_2)  -80 \hbar^4 \bigr) s^2
\end{split}
\end{equation}

In ref \cite{Das01}  the unitary representation of the algebra (\ref{eq:QLTS}) is studied.  In the case of $\gamma= 0,\beta=0$ and $\epsilon \neq 0$, the eigenvalues of the operator $A$ are given by the formula
\begin{equation}\label{eq:Ax}
A(x)= \sqrt{\epsilon} (x + u)
\end{equation}
where $x=0,\,1,\,2,\ldots$. The energy eigenvalues of the operator $H$ with degeneracy equal to $p+1$, where $p=0,\,1,\,2,\ldots$ are determined by solving the system of equations
\begin{equation}\label{eq:Phirel}
\Phi(0,u)=0,\; \Phi(p+1,u)=0.\; \mbox{ and }\; \Phi(x,u)>0 \; \mbox{where }\,
x=1,\,2,\,\ldots, p
\end{equation}
The structure function is given by the next relation,
\begin{equation}
\begin{split}
& \hspace{5em}  \Phi^q(x) = \frac{1}{4} \left ( -\frac{K}{\epsilon} -\frac{z}{\sqrt{\epsilon}} -\frac{\delta}{\sqrt{\epsilon}}\frac{\zeta}{\epsilon} + \frac{\zeta^2}{\epsilon^2}  \right ) - \\ - & \frac{1}{12}\left ( 3 d - a \sqrt{\epsilon} - 3 \alpha \ \frac{\delta}{\sqrt{\epsilon}} +3 \left ( \frac{\delta}{\sqrt{\epsilon}}\right )^2 - 6 \frac{z}{\sqrt{\epsilon}} + 6 \alpha \frac{\zeta}{\epsilon}- 6 \frac{\delta}{\sqrt{\epsilon}}\frac{\zeta}{\epsilon}\right ) (x+u) +\\ + & \frac{1}{4}\left (\alpha^2 + d - a \sqrt{\epsilon} - 3 \alpha \frac{\delta}{\sqrt{\epsilon}}+\left ( \frac{\delta}{\sqrt{\epsilon}}\right )^2 + 2 \alpha \frac{\zeta}{\epsilon}\right) (x+u)^2 - \\ - & \frac{1}{6} \left ( 3 \alpha^2 - a \sqrt{\epsilon} -3 \alpha \frac{\delta}{\sqrt{\epsilon}}  \right) (x+u)^3 + \frac{1}{4} \alpha^2 (x+u)^4
\end{split}
\end{equation}

In case of the \emph{q-system} the structure function is written
\begin{equation}\label{eq:qsyststrucfun}
\begin{split}
\Phi^q(x) = \frac{1}{\nu^4}\Bigl ( \hbar \nu \bigl ( 2(x+u) - 1 + \nu_2  \bigr ) - s \Bigr )\Bigl ( \hbar \nu \bigl ( 2(x+u) - 1 - \nu_2  \bigr ) -s \Bigr ) \\ \Bigl ( \hbar \nu \bigl ( 2(x+u) - 1 + \nu_1  \bigr ) -\mathcal{H}_2 \Bigr ) \Bigl (  \hbar \nu \bigl ( 2(x+u) - 1 - \nu_1  \bigr ) -\mathcal{H}_2 \Bigr )
\end{split}
\end{equation}
where
\[
\lambda = \frac{\nu^2}{2}, \; \lambda_i = \frac{\hbar^2 (\nu_i^2 - \frac{1}{4})}{2}
\]

In \cite{Das01} this method is used for the calculation of energy eigenvalues for the two dimensional superintegrable systems on the plane while in \cite{TanDas11b} the method have been extended to cover the case of three dimensional Kepler-Coulomb system. This quadratic, cubic and generally polynomial algebras is the subject of current investigations.  The cubic extension of the above algebra for superintegrable systems with an integral of motion cubic in momenta is studied in a series of papers by I. Marquette and P. Winternitz  \cite{MarWin07JMP48,MarWin08JPA41}, by I. Marquette \cite{Mar09JMP50a,Mar09JMP50b,Mar09JMP50c}. The case of higher order integrals of motion can be found in ref \cite{Mar10JPA43}, the case of one dimensional Position-Dependent Mass Schroedinger equation in \cite{Quesne07_SIGMA}. These methods are applied to three dimensional MICZ-Kepler system in \cite{Mar10JMP51}

\section{The quantum form of the `c' system. }\label{sec:eigencsys}

In contrary to the quantum form of the \emph{q-system} the quantum analog of the \emph{c-system} is not so obvious. This is because of the integral $B_1$ which is a fourth order polynomial in momenta. Specifically, the integrals which are constitute the quantum \emph{c-system} are written as follows
\[
\mathcal{H}_1 = \frac{1}{2} \left( p_{x_1}^2 +p_{x_2}^2 \right ) + \kappa \left( x_1^2 + 4 x_2^2 \right ) + \frac{\kappa_1}{x_1^2} + \frac{\kappa_2}{x_2^2} + r
\]
\begin{equation} \label{eq:quantumCsystem}
A_1 = \frac{1}{2} p_{x_1}^2 +  \kappa x_1^2 + \frac{\kappa_1}{x_1^2} + r
\end{equation}
\[
\begin{split}
B_1 = \left ( \frac{1}{2} \{p_{x_1},J_1\}\right )^2 + \frac{\kappa_2}{x_2^2} \{x_1^2,p_{x_1}^2\} & + \frac{1}{2} \left \{ Q, \frac{4 x_2 (\kappa_1-\kappa x_1^4)}{x_1^4} \right \}+ \\ &   + \frac{\hbar^2}{2} \left ( \frac{2 \kappa_2}{x_2^2} - 4 \kappa x_1^2 + \frac{4 \kappa_1 (x_1^2+8 x_2^2)}{x_1^4}  \right )
\end{split}
\]
where
\[
p_{x_1} = - i \hbar \partial_{x_1}, \; p_{x_2} = - i \hbar \partial_{x_2}, \; J_1 = - i \hbar (x_1 \partial_{x_2}-x_2 \partial_{x_1})
\]
and
\[
Q = -\frac{1}{2} \{x_1^3, p_{x_1} p_{x_2}\} + \frac{x_2}{2} \{x_1^2,p_{x_1}^2\} + \kappa_1 x_2 - \kappa x_2 x_1^4
\]

This system, as in classical mechanics case, forms a cubic triple algebraic structure which, in general, is given by the  following relations
\begin{equation}\label{eq:AABcubgen}
[A,[A,B]] = a A^3 + \alpha A^2 + \beta \{A,B\} + \gamma A + \delta B + \epsilon
\end{equation}
\begin{equation}\label{eq:BABcubgen}
[B,[A,B]] = \mu A^3 + \nu A^2 - \beta B^2 - a \{A,A,B\} - \alpha \{A,B\} + \xi A -\gamma B + \zeta
\end{equation}
where $\{A,A,B\}= A^2 B + A B A + B A^2$ and the structure constants are
\[
\alpha = \alpha_1 H + \alpha_0, \; \gamma = \gamma_2 H^2 + \gamma_1 H + \gamma_0, \;  \delta = \delta_1 H + \delta_0
\]
\[
\epsilon = \epsilon_3 H^3 + \epsilon_2 H^2 +\epsilon_1 H + \epsilon_0, \; \mu = \mu_1 H + \mu_0, \; \nu = \nu_2 H^2 + \nu_1 H + \nu_0
\]
\[
\xi = \xi_3 H^3 + \xi_2 H^2 + \xi_1 H +\xi_0, \; \zeta = \zeta_4 H^4 + \zeta_3 H^3 + \zeta_2 H^2 + \zeta_1 H + \zeta_0
\]
where $a, \beta, \alpha_i, \gamma_i, \delta_i, \epsilon_i, \mu_i, \nu_i, \xi_i, \zeta_i $ arbitrary constants and $H$ the Hamiltonian of the system.

The general formula for the Casimir operator is written
\[
\begin{split}
K = & [A,B]^2  -\frac{a}{2}\{A,A,A,B\} - \frac{2 \alpha}{3} \{A,A,B\} -\frac{2 \beta}{3} \{A,B,B\} + \left ( \frac{2 \beta^2}{3} -\delta \right ) B^2  + \\ & + \frac{1}{6} \left ( 2 \alpha \beta - 6 \gamma + a (\beta^2 - 3 \delta) \right ) \{A,B\} + \frac{1}{6}(4 \beta \gamma - 2\alpha \delta + a \beta \delta -12 \epsilon) B+\\ &+ \frac{1}{2}(3 a^2 + \mu) A^4  + \left ( 2 a \alpha -\frac{a^2 \beta}{2} + \beta \mu + \frac{2 \nu}{3} \right ) A^3 +\\ & +\frac{1}{6} (4 \alpha^2 -  a \alpha \beta + 6 a \gamma + a^2 (\beta^2-3 \delta) -\beta^2 \mu + 3\delta \mu + 4 \beta \nu + 6 \xi) A^2 \\ &  + \frac{1}{6} (4 \alpha \gamma + a \beta \gamma - 2 a \alpha \delta + a^2 \beta \delta + 12 \zeta - \beta \delta \mu + 2 \delta \nu + 2 \beta \xi) A
\end{split}
\]
where $\{A,A,A,B\} = A^3 B + A^2 B A + A B A^2 + B A^3.$

In case of the \emph{c-system} the non-zero commutators are
\begin{equation}\label{eq:AABGenQuanCsyst}
\begin{split}
[A_1,[A_1,B_1]] & =  16 \hbar^2 (A_1-r)^3 -16 \hbar^2 (\mathcal{H}_1 - r)(A_1-r)^2 + 32 \hbar^2 \kappa B_1 -\\ & - 8 (8\hbar^2 \kappa \kappa_1  - 11 \hbar^4 \kappa)(A_1-r)+ 8 ( 8 \hbar^2 \kappa \kappa_1 -3 \hbar^4 \kappa) (\mathcal{H}_1-r)
\end{split}
\end{equation}
\begin{equation}\label{eq:BABGenQuanCsyst}
\begin{split}
 [B_1,[A_1, B_1]] = -16 \hbar^2 \{(A_1-r),(A_1-r),B_1\} + 16 \hbar^2 (\mathcal{H}_1-r) \{(A_1-r),B_1\} - \\ - 32 (8 \hbar^2 \kappa_2 - 11 \hbar^4) (A_1-r)^3 + 8 ( 8 \hbar^2 \kappa \kappa_1 - 11 \hbar^4 \kappa)(A_1-r) B_1 + \\ + 16 \kappa (64 \hbar^2 \kappa_1 \kappa_2 - 40 \hbar^4 \kappa_1 -88 \hbar^4 \kappa_2 + 31 \hbar^6) (A_1-r)-  \\ - 352 \hbar^4 (\mathcal{H}_1-r)(A_1-r)^2 + 64 \hbar^4 (\mathcal{H}_1-r)^2(A_1-r)+ \\ + 48 \kappa (8 \hbar^4 \kappa_1 -3 \hbar^6) (\mathcal{H}_1-r) +
 \end{split}
\end{equation}
and the structure constants, based on the general form (\ref{eq:AABcubgen}), (\ref{eq:BABcubgen}) are
\[
a = 16 \hbar^2, \; \alpha = -16 \hbar^2 (\mathcal{H}_1-r)-48 \hbar^2 r,
\]
\[
  \gamma = 32 \hbar^2 (\mathcal{H}_1-r) r -64 \hbar^2 \kappa \kappa_1 + 48 \hbar^2 r^2 + 88 \hbar^4 \kappa ,
\]
\[
\; \beta = 0,\;\delta = 32 \hbar^2 \kappa, \;\mu = -256 \hbar^2 \kappa_2 + 352 \hbar^4,
\]
\[
\begin{split}
 \epsilon = -8 (2 \hbar^2 r^2 - 8 \hbar^2 \kappa \kappa_1 + 3 \hbar^4 \kappa )(\mathcal{H}_1-r) -\\- 16 \hbar^2 r^3 + 64 \hbar^2 \kappa \kappa_1 r - 88 \hbar^4 \kappa r ,
\end{split}
\]
\begin{equation}\label{eq:CubStrucConstGen}
\nu = -352 \hbar^4 (\mathcal{H}_1-r) + 96 r (8 \hbar^2 \kappa_2 - 11 \hbar^4 )
\end{equation}
\[
\begin{split}
\xi = 64  \hbar^4 (\mathcal{H}_1-r)^2 + 704 \hbar^4 r (\mathcal{H}_1-r) -96 (8 \hbar^2 \kappa_2 - 11 \hbar^4) r^2 + \\ + 16 \kappa (64 \hbar^2 \kappa_1 \kappa_2 - 8 \hbar^4 (5 \kappa_1+ 11 \kappa_2) + 31 \hbar^6  )
\end{split}
\]
\[
\begin{split}
\zeta = -64 \hbar^4 (\mathcal{H}_1-r)^2 r + 16 (24 \hbar^4 \kappa \kappa_1 - 22 \hbar^4 r^2 - 9 \hbar^6 \kappa)(\mathcal{H}_1-r)+\\ + 32 (8 \hbar^2 \kappa_2 - 11 \hbar^4) r^3  - 16 \kappa (64 \hbar^2 \kappa_1 \kappa_2 - 8 \hbar^4 (5 \kappa_1+ 11 \kappa_2) + 31 \hbar^6 ) r
\end{split}
\]
Therefore the Casimir operator is written
\begin{equation}
\begin{split}\label{eq:CsystCasim}
K = - \frac{64}{3} \biggl(11 \hbar^4 r^2 - 8 \hbar^4 \kappa \kappa_1 + 3 \hbar^6 \kappa \biggr ) (\mathcal{H}_1-r)^2 -\frac{32}{3} \biggl(70 \hbar^4 r^3 - \\ -136 \hbar^4 \kappa \kappa_1 r + 211 \hbar^6 \kappa r \biggr )(\mathcal{H}_1-r) + 16 (8 \hbar^2 \kappa_2 - 35 \hbar^4) r^4 - \\ - 16 \kappa \bigl(64 \hbar^2 \kappa_1 \kappa_2 - 8 \hbar^4 (13 \kappa_1 + 43 \kappa_2) +215 \hbar^6 \bigr) r^2 + \\ + 4 \kappa^2 \hbar^2 (8  \kappa_1 - 35 \hbar^2)(8  \kappa_1 - 3 \hbar^2)(8 \kappa_2 - 3 \hbar^2)
\end{split}
\end{equation}

Following the method described in \cite{Das01} we can achieve a finite representation of the cubic algebra in order to calculate the energy eigenvalues. Precisely, we construct a realization of the cubic algebra in terms of a deformed
oscillator algebra $\{b,b^\dag,\mathcal{N}\} $ which satisfies the relation
\begin{equation}\label{eq:defoscilalgebra}
[\mathcal{N},b^\dag]=b^\dag, \quad [\mathcal{N},b]=-b, \quad b^\dag b = \Phi(\mathcal{N}),\quad b b^\dag =\Phi(\mathcal{N}+1)
\end{equation}
where  $\Phi(x)$ a real valued function.

Let us consider the following realization
\begin{equation} \label{eq:realizcubAB}
A = A(\mathcal{N}), \; B = b(\mathcal{N}) + b^\dag(\mathcal{N}) \rho(\mathcal{N})+ \rho(\mathcal{N}) b(\mathcal{N})
\end{equation}
where the functions $A(x), b(x), \rho(x)$ will be determined by the cubic algebra. The relation $[A,B]=C$ implies
\[
C= b^\dag(\mathcal{N}) \Delta A(\mathcal{N}) \rho(\mathcal{N}) - \rho(\mathcal{N}) \Delta A(\mathcal{N}) b(\mathcal{N})
\]
where
\[
\Delta A(\mathcal{N}) =  A(\mathcal{N}+1) - A(\mathcal{N})
\]
using the relations (\ref{eq:AABcubgen}), (\ref{eq:realizcubAB}) we get
\[
\begin{split}
[A,C] & = [A(\mathcal{N}), b^\dag(\mathcal{N}) \Delta A(\mathcal{N}) \rho(\mathcal{N}) - \rho(\mathcal{N}) \Delta A(\mathcal{N}) b(\mathcal{N})] \\ & = b^\dag(\mathcal{N}) \left (\Delta A(\mathcal{N}) \right )^2 \rho(\mathcal{N}) + \rho(\mathcal{N}) \left (\Delta A(\mathcal{N}) \right )^2 b(\mathcal{N}) \\ & = a A^3 + \alpha A^2 + \beta \{A,B\} + \gamma A + \delta B + \epsilon \\ & = b^\dag(\mathcal{N}) \Bigl ( \beta \bigl ( A(\mathcal{N}+1)+A(\mathcal{N}) \bigr )+\delta  \Bigr ) \rho(\mathcal{N}) + \\ & \hspace{5em} + \rho(\mathcal{N}) \Bigl (  \beta \bigl ( A(\mathcal{N}+1)+A(\mathcal{N}) \bigr )+\delta \Bigr ) b(\mathcal{N}) + \\ & +  a A^3(\mathcal{N}) + \alpha A^2(\mathcal{N}) + 2 \beta A(\mathcal{N}) b(\mathcal{N}) + \gamma A(\mathcal{N}) + \delta b(\mathcal{N}) + \epsilon
\end{split}
\]
therefore we have the following relations
\begin{equation}\label{eq:firstrel}
\Bigl ( \Delta A(\mathcal{N}) \Bigr )^2 = \beta \Bigl ( A(\mathcal{N}+1) + A(\mathcal{N}) \Bigr ) + \delta
\end{equation}
\begin{equation}\label{eq:secondrel}
 a A^3(\mathcal{N}) + \alpha A^2(\mathcal{N}) + 2 \beta A(\mathcal{N}) b(\mathcal{N}) + \gamma A(\mathcal{N}) + \delta b(\mathcal{N}) + \epsilon = 0
\end{equation}
while the function $\rho(\mathcal{N})$ can be arbitrarily determined. In fact this function can be fixed, in order
to have a polynomial structure function $\Phi(x)$ for the deformed oscillator algebra (\ref{eq:defoscilalgebra}).
The solutions of equation (\ref{eq:firstrel}) depend on the value of the parameter $\beta$, while the function
$b(x)$ is uniquely determined by Eq. (\ref{eq:secondrel}) (provided that almost one of the parameters $\beta$ or $\epsilon$ is not
zero). At this stage, the cases $\beta=0$ or $\beta\neq 0$ should be treated separately. We can see the following.

\noindent {\textbf{ Case I. $\beta \neq 0$}}

In this case the solutions of Eqs. (\ref{eq:firstrel}) and (\ref{eq:secondrel}) are given by
\[
A(\mathcal{N}) = \frac{\beta}{2} \left ( \bigl( \mathcal{N}+u \bigr )^2 - \frac{1}{4} - \frac{\delta}{\beta^2} \right )
\]
\[
\begin{split}
b(\mathcal{N})  =  \frac{8 \alpha \beta \Bigl ( \bigl (1-4 (\mathcal{N}+u)^2 \bigr)  \beta^2 + 4 \delta \Bigr )^2}{128 (1- 4 (\mathcal{N}+u)^2) \beta^5}  - \frac{a \Bigl ( \bigl (1-4 (\mathcal{N}+u)^2 \bigr) \beta^2 + 4 \delta \Bigr )^3}{128 (1- 4 (\mathcal{N}+u)^2) \beta^5} - \\   - \frac{64 \beta^2 ( (1- 4 (\mathcal{N}+u)^2) \beta^2 \gamma + 4 \gamma \delta - 8 \beta \epsilon  )}{128 (1- 4 (\mathcal{N}+u)^2) \beta^5}
\end{split}
\]

\noindent {\textbf{ Case II. $\beta = 0, \; \delta \neq 0$}}

In this case the solutions of Eqs. (\ref{eq:firstrel}) and (\ref{eq:secondrel}) are given by
\begin{equation}\label{eq:ACubRepr}
A(\mathcal{N}) = \sqrt{\delta} (\mathcal{N}+ u)
\end{equation}
\begin{equation}
\begin{split}
b(\mathcal{N}) =  - \frac{\gamma (\mathcal{N}+u)}{2 \beta (\mathcal{N}+u)+\sqrt{\delta}} - \frac{\alpha \sqrt{\delta}(\mathcal{N}+u)^2}{2 \beta (\mathcal{N}+u)+\sqrt{\delta}} - & \frac{a \delta (\mathcal{N}+u)^3}{2 \beta (\mathcal{N}+u)+\sqrt{\delta}} - \\  & - \frac{\epsilon}{2 \beta \sqrt{\delta} (\mathcal{N}+u)+ \delta }
\end{split}
\end{equation}

Using the above definitions (\ref{eq:realizcubAB})  for $A(\mathcal{N})$ and $b(\mathcal(N))$ the relation (\ref{eq:BABcubgen}) leads to the following expression
\begin{equation}\label{eq:realBC}
\begin{split}
[B,C] & =  2 \Phi(\mathcal{N}+1) \left ( \Delta A(\mathcal{N}) + \frac{\beta}{2} \right ) \rho(\mathcal{N}) - 2 \Phi(\mathcal{N}) \left ( \Delta(\mathcal{N}-1)-\frac{\beta}{2} \right ) \rho(\mathcal{N}-1) \\ & = \mu A^3(\mathcal{N}) + \nu A^2(\mathcal{N}) - 3 a A^2(\mathcal{N}) b(\mathcal{N}) - \beta b^2(\mathcal{N}) - 2\alpha A(\mathcal{N}) b(\mathcal{N}) + \\ & \hspace{15em} + \xi A(\mathcal{N}) -\gamma b(\mathcal{N}) + \zeta
\end{split}
\end{equation}
the Casimir operator is now realized as
\begin{equation}\label{eq:realCasimir}
\begin{split}
K  & = \Phi(\mathcal{N}+1) \left ( \frac{2 \beta^2}{3} -\delta - 2 \beta A(\mathcal{N}) - \Delta A^2(\mathcal{N}) \right ) \rho(\mathcal{N}) + \\ & + \Phi(\mathcal{N}) \left ( \frac{2 \beta^2}{3} -\delta - 2 \beta A(\mathcal{N}) - \Delta A^2(\mathcal{N}-1) \right ) \rho(\mathcal{N}-1)  -2 a A^3(\mathcal{N}) b(\mathcal{N})-  \\ &  - 2 \alpha A^2(\mathcal{N}) b(\mathcal{N})-2 \beta A(\mathcal{N}) b^2(\mathcal{N}) + \frac{1}{3} \left ( 2 \alpha \beta - 6 \gamma + a (\beta^2 - 3 \delta) \right ) A(\mathcal{N}) b(\mathcal{N})  + \\ & + \left ( \frac{2 \beta^2}{3} -\delta \right ) b^2(\mathcal{N})  + \left ( 2 a \alpha -\frac{a^2 \beta}{2} + \beta \mu + \frac{2 \nu}{3} \right ) A^3(\mathcal{N}) +\\  & +\frac{1}{6} (4 \alpha^2 - 3 a \alpha \beta + 6 a \gamma + a^2 (\beta^2-3 \delta) -\beta^2 \mu + 3\delta \mu + 4 \beta \nu + 6 \xi) A^2(\mathcal{N}) + \\  &  + \frac{1}{6}(4 \beta \gamma - 2\alpha \delta + a \beta \delta -12 \epsilon) b(\mathcal{N}) + \frac{1}{2}(3 a^2 + \mu) A^4(\mathcal{N})   + \\  & + \frac{1}{6} (4 \alpha \gamma + a \beta \gamma - 2 a \alpha \delta + a^2 \beta \delta + 12 \zeta - \beta \delta \mu + 2 \delta \nu + 2 \beta \xi) A(\mathcal{N})
\end{split}
\end{equation}
Equations (\ref{eq:realBC}) and (\ref{eq:realCasimir}) are linear functions of the expressions $\Phi(\mathcal{N})$ and $\Phi(\mathcal{N}+1)$ then the function $\Phi(\mathcal{N})$ can be determined, if the function $\rho(\mathcal{N})$ is given. The function $\Phi(\mathcal{N})$  depends on two parameters $u$ and $K$ and it is given by the following formulas

\noindent {\textbf{ Case I. $\beta \neq 0$}}
\[
\begin{split}
\rho(\mathcal{N}) =  \frac{1}{\bigl ( 12 (\mathcal{N}+u+1)^4 + 6 (\mathcal{N}+u+1)^3 -10 (\mathcal{N}+u+1)^2 -5 (\mathcal{N}+u+1)\bigr )} \cdot \\ \cdot \frac{1}{2^{17} \beta^{10} \bigl (  2 (\mathcal{N}+u) +1 \bigr )^2}
\end{split}
\]
The function $\Phi(\mathcal{N})$ is a polynomial of order 14 and is given in the appendix A.

\noindent {\textbf{ Case II. $\beta = 0, \; \delta \neq 0$}}

\[
\rho(\mathcal{N}) = \frac{1}{24 \delta^2}
\]

\begin{equation}\label{eq:CubStrucGen}
\begin{split}
\Phi^c(\mathcal{N}) & = -6 \delta K + 6 a^2 \delta^3 (\mathcal{N}+u)^6 +  ( 12 a \alpha \delta^{5/2}-18 a^2 \delta^3  ) (\mathcal{N}+u)^5 + \\ & + 3 \delta^2 ( 5 a^2 \delta +2 \alpha ^2-10 \alpha  a \sqrt{\delta }+4 a \gamma +\delta  \mu) (\mathcal{N}+u)^4+2 \delta^{3/2} \Bigl( -6 \alpha ^2 \sqrt{\delta } + \\ & +2 \alpha  (5 a \delta +3 \gamma )+6 a (\epsilon -2 \gamma  \sqrt{\delta })-3 \delta ^{3/2} \mu +2 \delta  \nu \Bigr) (\mathcal{N}+u)^3 + \\ & + 3 \delta (-a^2 \delta ^2+2 \alpha ^2 \delta -6 \alpha  \gamma  \sqrt{\delta }+4 a \gamma  \delta -6 a \sqrt{\delta } \epsilon +4 \alpha  \epsilon +2 \gamma ^2-2
   \delta ^{3/2} \nu + \\ & +\delta ^2 \mu +2 \delta  \xi) (\mathcal{N}+u)^2 + 2 \sqrt{\delta}(a \delta  (3 \epsilon -\alpha  \delta )+3 \gamma  (\alpha  \delta +2 \epsilon )-6 \alpha  \sqrt{\delta } \epsilon - \\ & -3 \gamma ^2 \sqrt{\delta }-3 \delta
   ^{3/2} \xi +\delta ^2 \nu +6 \delta  \zeta)(\mathcal{N}+u) + 2 \alpha  \delta  \epsilon -6 \gamma  \sqrt{\delta } \epsilon -6 \delta ^{3/2} \zeta +6 \epsilon ^2
\end{split}
\end{equation}

In case of the \emph{c-system} the above structure function is given by the next formula
\begin{equation}\label{eq:csyststrucfun}
\begin{split}
 \Phi^c(y)   = & 3 \cdot 2^{9} \hbar^{4}  \Bigl(   \hbar \nu \bigl( 4(y+v)-1 -\nu_1 \bigr ) - r \Bigr ) \Bigl( \hbar \nu \bigl( 4(y+v)-3 -\nu_1 \bigr ) - r \Bigr) \\ & \Bigl(  \hbar \nu \bigl ( 4(y+v)-1 + \nu_1 \bigr )  - r \Bigr) \Bigl(  \hbar \nu \bigr ( 4(y+v)-3 + \nu_1 \bigr) - r \Bigr) \\ & \Bigl(  2  \hbar \nu \bigl(2(y+v)-1 + \nu_2 \bigr) - \mathcal{H}_1\Bigr) \Bigl( 2  \hbar \nu\bigl(2(y+v)-1 - \nu_2 \bigr) - \mathcal{H}_1 \Bigr)
\end{split}
\end{equation}

where
\[
\kappa = \frac{\nu^2}{2}, \; \kappa_i = \frac{\hbar^2 (\nu_i^2 - \frac{1}{4})}{2}
\]

\section{Generalized anisotropic oscillator systems \\ through Hamiltonian Chains}

In general the generalized three dimensional anisotropic oscillator potential is the classical anisotropic oscillator potential in addition to the so-called Smo\-ro\-din\-ski-Winternitz terms. Namely, these systems are characterized by a Hamiltonian function of the following form
\begin{equation}
H= \frac{1}{2} ( p_x^2+ p_y^2 +p_z^2)+ k (a_1^2 x^2+ a_2^2 y^2 + a_3^2 z^2) + \frac{k_1}{x^2} +  \frac{k_2}{y^2}+  \frac{k_3}{z^2}
\end{equation}
Since these systems can be classified using the constants $a_i$  in the next sections we shall use the notation $(a_1,a_2,a_3)$ for all three dimensional generalized anisotropic oscillator potentials.

\subsection{Generalized anisotropic oscillator $(1,1,2)$ system}

Let us consider  the following  sequence of Hamiltonians
\[
H_1 = \frac{1}{2} p_x^2 + k x^2 + \frac{k_1}{x^2}
\]
\begin{equation}\label{eq:syst112}
H_2 = \frac{1}{2} ( p_x^2+ p_y^2 )+ k (x^2+y^2) + \frac{k_1}{x^2} +  \frac{k_2}{y^2}
\end{equation}
\[
H_3 = \frac{1}{2} ( p_x^2+ p_y^2 +p_z^2)+ k (x^2+y^2 + 4 z^2) + \frac{k_1}{x^2} +  \frac{k_2}{y^2}+  \frac{k_3}{z^2}
\]
From now on these sequences will be refereed as \emph{Hamiltonian chains}. The coefficients $a_i$ of the last Hamiltonian characterize the chain namely the above sequence will be refereed as $(1,1,2)$ anisotropic oscillator chain.  The above chain can be split into two two-dimensional superintegrable systems. Specifically, setting to the \emph{q-system}
\begin{equation}\label{eq:subQ112}
q_1 = y, q_2 = x, \lambda = k, \lambda_1 = k_2, \lambda_2 = k_1 , s = 0
\end{equation}
we obtain $\mathcal{H}_2 \equiv H_2$ and $A_2 \equiv H_1.$

Similarly, setting to the \emph{c-system}
\begin{equation}\label{eq:subC112}
x_1 = y,x_2 = z, \kappa = k, \kappa_1 = k_2, \kappa_2 = k_3 , r = A_2 \equiv H_1
\end{equation}
we have $\mathcal{H}_1 \equiv H_3, A_1 \equiv H_2.$ Therefore the hamiltonian chain (\ref{eq:syst112}) is characterized by the existence of one \emph{q-system} and one \emph{c-system}. From the relations (\ref{eq:generatcubgener}), (\ref{eq:generatquadrgener}) it is straightforward to calculate the generating functions of the corresponding triple cubic and quadratic  Poisson algebras (\ref{eq:cubicclassalgebra}), (\ref{eq:quadraclassalgebra})

This \emph{Hamiltonian chain} can be presented by the following diagram
\begin{equation}\label{eq:chain112}
 \xymatrix{      &       &        &H_1 \ar@{--}[lld]\ar@{--}[ld]\ar@{--}[rd]\ar@{--}[rrd]
\\ & B_2  \ar@{--}[r] &H_2 \ar@{--}[r]& \ar@{--}[r] & H_3 \ar@{--}[r] & B_1}
 \end{equation}
where the dashed line joining two integrals  means that the corresponding commutator is zero.

The quantum form of the two systems composing the \emph{Hamiltonian chain} can be derived from the expressions (\ref{eq:quantumqsystem}), (\ref{eq:quantumCsystem}). Furthermore, the form of their quantum algebras as well as the form of the Casimir operator are straightforward calculated by the relations (\ref{eq:AABGenQuanQsyst})-(\ref{eq:qsystCasim}), (\ref{eq:AABGenQuanCsyst})-(\ref{eq:CsystCasim}) using the above substitutions (\ref{eq:subQ112}), (\ref{eq:subC112}) for the \emph{q-system} and \emph{c-system} respectively.

We are interested to calculate the energy eigenvalues of the operators $H_i$.
Setting
\[
k=\frac{\mu^2}{2}, \; k_i = \frac{\hbar^2 (\mu_i^2 - \frac{1}{4})}{2}
\]
the structure function $\Phi^q( y)$ for the \emph{q-system} is written
\begin{equation}
\begin{split}
\begin{split}
\Phi^q( y) = \frac{\hbar^2}{\mu^2}    ( 2  y - 1 + \mu_1   )     ( 2  y - 1 - \mu_1   ) \bigl ( H_2 - \hbar \mu  ( 2  y - 1 + \mu_2  ) \bigr ) \\ \bigl ( H_2 - \hbar \mu  ( 2  y - 1 - \mu_2   ) \bigr )
\end{split}
\end{split}
\end{equation}
where $ y = x + u.$ Furthermore, the eigenvalues of the operator $H_1$  is given using the relation (\ref{eq:Ax})
\[
H_1= \sqrt{\epsilon} y = 2 \hbar \mu  y
\]
Setting
\[
h_1 = \frac{H_1}{2 \hbar \mu} =  y, \quad h_2 = \frac{H_2}{2 \hbar \mu}
\]
the structure function is written
\[
\begin{split}
\Phi^q = 16 \hbar^4  \left (  \left(h_1 - \frac{1}{2}\right)^2 - \left(\frac{\mu_1}{2}\right)^2   \right )  \left ( \left(h_2-h_1+\frac{1}{2}\right)^2 - \left(\frac{\mu_2}{2}^2 \right) \right )
\end{split}
\]
The function $\Phi^q(x)$ is a real valued, positive function, which for a finite representation must satisfies the boundary conditions $\Phi^q(0)=\Phi^q(q+1)=0$ where $q$ any positive integer.
The system of the boundary conditions lead us to the following expressions for the eigenvalues of $h_1$ and $h_2$
\begin{equation}
h_1 = m + u, \quad m=0,\ldots,q,
\end{equation}
where
\[
u=\frac{1}{2} + \frac{\mu_1}{2}
\]
\begin{equation}\label{eq:h2q112}
h_2 = q +1 +\frac{\mu_1}{2} + \frac{\mu_2}{2}
\end{equation}

The structure function is now written
\[
\Phi^q(m) = 16 \hbar^4 m (q+1-m)(m +\mu_1)(q+1-m +\mu_2)
\]
Also, since $\Phi^q(m) > 0$  must be $ \mu_1 > -1, \mu_2 > -1. $

Similarly, the structure function for the \emph{c-system} is given by the following relation
\begin{equation}
\begin{split}
 \Phi^c(x)   = & 3 \cdot 2^{9} \hbar^{4}  \bigl( H_1 -  \hbar \mu ( 4 x-1 -\mu_2  )  \bigr ) \bigl( H_1 - \hbar \mu ( 4 x-3 -\mu_2  )  \bigr) \\ & \bigl( H_1 -  \hbar \mu  ( 4 x-1 + \mu_2  )  \bigr) \Bigl( H_1 -  \hbar \mu  ( 4 x-3 + \mu_2 )  \bigr) \\ & \bigl( H_3 - 2  \hbar \mu (2 x-1 + \mu_3 ) \bigr) \bigl(H_3 - 2  \hbar \mu (2 x-1 - \mu_3 )  \bigr)
\end{split}
\end{equation}
and the eigenvalues of the operator $H_2$ is given by the relation (\ref{eq:ACubRepr})
\[
H_2= \sqrt{\delta} x = 4 \hbar \mu x
\]
where $x=y+v$. Setting
\[
h_1 = \frac{H_1}{2 \hbar \mu}, \quad h_2 = \frac{H_2}{2 \hbar \mu} = 2 x, \quad h_3 = \frac{H_3}{2 \hbar \mu}
\]
the function $\Phi^c(x)$ is written
\[
\begin{split}
 \Phi^c  & =  3 \cdot 2^{15} \hbar^{10} \mu^6 \left ( \left(h_3-h_2+1\right)^2 - \mu_3^2  \right ) \\  &\left ( \left(h_2-h_1-\frac{1}{2}\right)^2 - \left(\frac{\mu_2}{2} \right)^2 \right )  \left ( \left(h_2-h_1-\frac{3}{2}\right)^2 - \left(\frac{\mu_2}{2} \right)^2 \right )
\end{split}
\]
The \emph{q-system} implies that the eigenvalues of the operator $h_1$ is given by the next formula
\[
h_1 = h_1^{min} + m ,\quad  \mbox{with} \quad h_1^{min}=u = \frac{1}{2} + \frac{\mu_1}{2}, \quad  m=0,\ldots,q.
\]
Finally the system of the boundary conditions $\Phi^c(m)=\Phi^c(p+1)=0$ where $\Phi^c(q)>0$ for all $ q=m+1,\ldots,p$  lead us to the following formulas for the eigenvalues of the operators $h_2, h_3$

\[
h_2 = v + q - m, \quad   v  = q + 1+ \frac{\mu_1}{2} + \frac{\mu_2}{2},  \quad m\leq q \leq p
\]

\[
h_3 = v + p - m  + \mu_3 = q + p - m + 1+\frac{\mu_1}{2} + \frac{\mu_2}{2} + \mu_3
\]
The structure function $\Phi^c(q)$ is now written
\[
\begin{split}
\Phi^c(q) = 3 \cdot 2^{16} \hbar^{10} \mu^6 (q-m) (p+1-q) (2 q- 1 - 2 m) (2 q - 2 m + \mu_2) \\ (2 q - 1 - 2 m + \mu_2)(p+1 - q + 2 \mu_3)
\end{split}
\]
Also, since $\Phi^c(q) > 0$  must be $ \mu_2 > -1, \mu_3 > -1/2. $

Consequently the final form  of the eigenvalues for the involving operators composing the chain is given as follows
\[
H_1 =  \hbar \mu (2 m + 1 + \mu_1), \quad H_2 = \hbar \mu  (4 q - 2 m + 2 + \mu_1 + \mu_2  )
\]
\[
H_3 = \hbar \mu \bigl(2 (q +  p -  m) + 4 + \mu_1 + \mu_2 + 2 \mu_3 \bigr)
\]
The corresponding structure functions are
\[
\begin{split}
\Phi^c(q) = 3 \cdot 2^{16} \hbar^{10} \mu^6 (q-m) (p+1-q) (2 q- 1 - 2 m) (2 q - 2 m + \mu_2) \\ (2 q - 1 - 2 m + \mu_2)(p+1 - q + 2 \mu_3)
\end{split}
\]
\[
\Phi^q(m) = 16 \hbar^4 m (q+1-m)(m +\mu_1)(q+1-m +\mu_2)
\]
\[
m,q,p \in \mathbb{N}, \quad 0 \leq m\leq q \leq p,\qquad \mu_1,\mu_2 > -1, \; \mu_3 > -\frac{1}{2}
\]

\subsection{Generalized anisotropic oscillator $(1,2,2)$ system}

The generalized anisotropic oscillator chain $(1,2,2)$ is given by the following sequence
\[
H_1 = \frac{1}{2} p_x^2 + k x^2 + \frac{k_1}{x^2}
\]
\begin{equation}\label{eq:chain122}
H_2 = \frac{1}{2} ( p_x^2+ p_y^2 )+ k (x^2 + 4 y^2) + \frac{k_1}{x^2} +  \frac{k_2}{y^2}
\end{equation}
\[
H_3 = \frac{1}{2} ( p_x^2+ p_y^2 +p_z^2)+ k (x^2+4 y^2 + 4 z^2) + \frac{k_1}{x^2} +  \frac{k_2}{y^2}+  \frac{k_3}{z^2}
\]
As in previous case this chain implies the existence of two two-dimensional superintegrable systems. Precisely, setting to the \emph{c-system}
\begin{equation}\label{eq:subC122}
x_1 = x,x_2 = y, \kappa = k, \kappa_1 = k_1, \kappa_2 = k_2 , r = 0
\end{equation}
we obtain $\mathcal{H}_1 \equiv H_2$ and $A_1 \equiv H_1$

Furthermore setting to the \emph{q-system}
\begin{equation}\label{eq:subQ122}
q_1 = z, q_2 = y, \lambda =4 k, \lambda_1 = k_3, \lambda_2 = k_2 , s = A_1 \equiv H_1
\end{equation}
we have $\mathcal{H}_2 \equiv H_3$ and $A_2 \equiv H_2.$
The relations (\ref{eq:generatcubgener}), (\ref{eq:generatquadrgener}) lead us to the generating functions of the corresponding triple cubic and quadratic  Poisson algebras (\ref{eq:cubicclassalgebra}), (\ref{eq:quadraclassalgebra}).

The quantum form of the two systems composing the $(1,2,2)$ \emph{Hamiltonian chain} can be derived from the expressions (\ref{eq:quantumqsystem}), (\ref{eq:quantumCsystem}). Furthermore, the form of their quantum algebras as well as the form of the Casimir operator are straightforward calculated by the relations (\ref{eq:AABGenQuanQsyst})-(\ref{eq:qsystCasim}), (\ref{eq:AABGenQuanCsyst})-(\ref{eq:CsystCasim}) using the above substitutions (\ref{eq:subC122}), (\ref{eq:subQ122}) for the \emph{c-system} and \emph{q-system} respectively.

As in $(1,1,2)$ chain  we are interested to calculate the energy eigenvalues of the operators $H_i$.
Setting
\[
k=\frac{\mu^2}{2}, \; k_i = \frac{\hbar^2 (\mu_i^2 - \frac{1}{4})}{2}
\]
the structure function $\Phi^c(x)$ for the \emph{c-system} is written
\[
\begin{split}
\Phi^c(x) = 3 \cdot 2^9 \hbar^8 \mu^4  ( 4 x - 3 - \mu_1  )  ( 4 x - 3 + \mu_1  )  ( 4 x - 1 + \mu_1   )  ( 4 x - 1 - \mu_1   ) \\ \bigl ( H_2 - 2 \hbar \mu (2 x - 1 - \mu_2 )  \bigr )  \bigl ( H_2 - 2 \hbar \mu (2 x - 1 + \mu_2  )  \bigr )
\end{split}
\]
and eigenvalues of the operator $H_1$ is given by the relation (\ref{eq:ACubRepr})
\[
H_1 =\sqrt{\delta} x =4 \hbar \mu x
\]
where $x=y+v$. Setting
\[
h_1 = \frac{H_1}{2 \hbar \mu}=2 x, \quad h_2 = \frac{H_2}{2 \hbar \mu}
\]
the function $\Phi^c(x)$ is written
\[
\begin{split}
\Phi_1 = 3 \cdot 2^{15} \hbar^{10} \mu^6 \left (  \left(h_1 - \frac{1}{2}\right)^2 - \left(\frac{\mu_1}{2}\right)^2   \right ) \left (  \left(h_1 - \frac{3}{2}\right)^2 - \left(\frac{\mu_1}{2}\right)^2   \right )\\ \left ( \left(h_2-h_1+1\right)^2 - \mu_2^2  \right )
\end{split}
\]

The system of the boundary conditions $\Phi^c(0)=\Phi^c(q+1)=0$  where $\Phi^c(m)>0$ for all $ m=1,\ldots,q$ lead us to the following expressions for the eigenvalues of $h_1$ and $h_2$

\[
h_1 = v + m, \quad m=0,\ldots,q, \qquad v = \frac{3}{2} + \frac{\mu_1}{2}
\]
\begin{equation}\label{eq:h2q122}
 h_2 = q +\frac{3}{2} +\frac{\mu_1}{2} + \mu_2
\end{equation}
The structure function is given as follows
\[
\Phi^c(m) = 3 \cdot 2^{15} \hbar^{10} \mu^6  m (m+1) (q+1-m) (m+1 + \mu_1) (m+\mu_1) (q+1-m + 2 \mu_2)
\]
Also, since $\Phi^c(m) > 0$  must be $ \mu_1 > -1, \mu_2 > -1/2. $

Similarly, the structure function for the \emph{q-system} is given by the following relation
\[
\begin{split}
\Phi^q(y)= \frac{1}{16 \mu^4} \bigl ( H_1 - 2 \hbar \mu ( 2 y - 1 - \mu_2 ) \bigr ) \bigl ( H_1 - 2 \hbar \mu ( 2 y - 1 + \mu_2 ) \bigr ) \\ \bigl( H_3 - 2 \hbar \mu  ( 2 y - 1 - \mu_3  )  \bigr )\bigl( H_3 - 2 \hbar \mu  ( 2 y - 1 + \mu_3  )  \bigr )
\end{split}
\]
where $ y = x + u.$ Moreover, the eigenvalues of the operator $H_2$  is given using the relation (\ref{eq:Ax})
\[
H_2= \sqrt{\epsilon} y = 4 \hbar \mu y
\]
Setting
\[
h_1 = \frac{H_1}{2 \hbar \mu}, \quad h_2 = \frac{H_2}{2 \hbar \mu} = 2 y, \quad h_3 = \frac{H_3}{2 \hbar \mu}
\]
the structure function is now written
\[
\begin{split}
\Phi^q =  \hbar^4  \left (  \left(h_2 - h_1 - 1 \right)^2 - \mu_2^2   \right )  \left ( \left( h_3-h_2+ 1 \right)^2 - \mu_3^2  \right )
\end{split}
\]
The \emph{c-system} implies that
\[
h_1 = \frac{3}{2} + \frac{\mu_1}{2} + m, \quad   m=0,\ldots,q
\]
Finally the system of the boundary conditions $\Phi^q(m)=\Phi^q(p+1)=0$ where $\Phi^q(q)>0$ for all $ q=m+1,\ldots,p$ lead us to the lead us to the following expressions for the eigenvalues of $h_2$ and $h_3$
\[
h_2 = u + q - m+1, \quad   u = q + \frac{3}{2}+ \frac{\mu_1}{2} + \mu_2,  \quad m\leq q \leq p
\]
\[
h_3 = q + p - m + \frac{5}{2}+ \frac{\mu_1}{2} + \mu_2 + \mu_3
\]
The structure function is now written
\[
\begin{split}
\Phi^q(q) =  4 \hbar^4 (q-m)(p+1-q)( q -  m +  \mu_2) (p+1 - q + 2 \mu_3)
\end{split}
\]
Also, since $\Phi^q(q) > 0$  must be $ \mu_2 > -1, \mu_3 > -1/2. $

The final form  of the eigenvalues for the involving operators composing the $(1,2,2)$ chain is given as follows
\[
H_1 =  \hbar \mu \left(2 (m+1) + 1  + \mu_1 \right), \quad H_2 = \hbar \mu  (4 q - 2 m + 5 + \mu_1 + 2 \mu_2  )
\]
\[
H_3 = \hbar \mu \bigl(2 (q +  p -  m) + 5 + \mu_1 + 2 \mu_2 + 2 \mu_3 \bigr)
\]
The corresponding structure functions are
\[
\Phi^c(m) = 3 \cdot 2^{15} \hbar^{10} \mu^6  m (m+1) (q+1-m) (m+1 + \mu_1) (m+\mu_1) (q+1-m + 2 \mu_2)
\]
\[
\Phi^q(q) =  4 \hbar^4 (q-m)(p+1-q)( q -  m +  \mu_2) (p+1 - q + 2 \mu_3)
\]
\[
m,q,p \in \mathbb{N}, \quad 0 \leq m\leq q \leq p,\qquad \mu_1 > -1, \; \mu_2, \mu_3 > -\frac{1}{2}
\]

\subsection{Generalized anisotropic oscillator $(1,2,4)$ system}

The generalized anisotropic oscillator chain $(1,2,4)$  is given by the following sequence

\[
H_1 = \frac{1}{2} p_x^2 + k x^2 + \frac{k_1}{x^2}
\]
\begin{equation}\label{eq:chain124}
H_2 = \frac{1}{2} ( p_x^2+ p_y^2 )+ k (x^2 + 4 y^2) + \frac{k_1}{x^2} +  \frac{k_2}{y^2}
\end{equation}
\[
H_3 = \frac{1}{2} ( p_x^2+ p_y^2 +p_z^2)+ k (x^2+4 y^2 + 16 z^2) + \frac{k_1}{x^2} +  \frac{k_2}{y^2}+  \frac{k_3}{z^2}
\]
This chain implies the existence of two \emph{c-systems} systems. Precisely, setting to the \emph{c-system}
\begin{equation}\label{eq:sub1C124}
x_1 = x,x_2 = y, \kappa = k, \kappa_1 = k_1, \kappa_2 = k_2 , r = 0
\end{equation}
we get $\mathcal{H}_1 \equiv H_2$ and $A_1 \equiv H_1.$

Furthermore, setting
\begin{equation}\label{eq:sub2C124}
x_1 = y,x_2 = z, \kappa = 4 k, \kappa_1 = k_2, \kappa_2 = k_3 , r = A_1 \equiv H_1
\end{equation}
we have $\mathcal{H}_1 \equiv H_3$ and $A_1 \equiv H_2.$
We shall refereed to these systems as first \emph{c-system} and second \emph{c-system}. The relation (\ref{eq:generatcubgener}) lead us to the generating functions of the corresponding triple cubic Poisson algebras (\ref{eq:cubicclassalgebra}).

The quantum form of the two \emph{c-systems} composing the $(1,2,4)$ \emph{Hamiltonian chain} can be derived from the expressions (\ref{eq:quantumCsystem}). Furthermore, the form of their quantum algebras as well as the form of the Casimir operator are straightforward calculated by the relations (\ref{eq:AABGenQuanCsyst})-(\ref{eq:CsystCasim}) using the above substitutions (\ref{eq:sub1C124}), (\ref{eq:sub2C124}) for the two \emph{c-systems}.

The calculation of the energy eigenvalues for the operators $H_i$ is started setting,
\[
k=\frac{\mu^2}{2}, \; k_i = \frac{\hbar^2 (\mu_i^2 - \frac{1}{4})}{2}
\]
the structure function $\Phi^c_1(x)$ for the first \emph{c-system} is written
\[
\begin{split}
\Phi^c_1(x) = 3 \cdot 2^9 \hbar^8 \mu^4  ( 4 x - 3 - \mu_1  )  ( 4 x - 3 + \mu_1  )  ( 4 x - 1 + \mu_1   )  ( 4 x - 1 - \mu_1   ) \\ \bigl ( H_2 - 2 \hbar \mu (2 x - 1 - \mu_2 )  \bigr )  \bigl ( H_2 - 2 \hbar \mu (2 x - 1 + \mu_2  )  \bigr )
\end{split}
\]
and eigenvalues of the operator $H_1$ is given by the relation (\ref{eq:ACubRepr})
\[
H_1 =\sqrt{\delta} x =4 \hbar \mu x
\]
where $x=y+v_1$. If we set
\[
h_1 = \frac{H_1}{2 \hbar \mu}=2 x, \quad h_2 = \frac{H_2}{2 \hbar \mu}
\]
The structure function is written
\[
\begin{split}
\Phi^c_1(x) = 3 \cdot 2^{15} \hbar^{10} \mu^6 \left (  \left(h_1 - \frac{1}{2}\right)^2 - \left(\frac{\mu_1}{2}\right)^2   \right ) \left (  \left(h_1 - \frac{3}{2}\right)^2 - \left(\frac{\mu_1}{2}\right)^2   \right )\\ \left ( \left(h_2-h_1+1\right)^2 - \mu_2^2  \right )
\end{split}
\]
The form of the operator $h_1$ eigenvalues is
\[
h_1 = v_1 + m, \quad m=0,\ldots,q
\]
The boundary conditions  $\Phi^c_1(0)=\Phi^c_1(q+1)=0$ where $q$ a positive integer as we have seen in $(1,2,2)$ chain lead us to the following choice of $v_1$
\[
v_1 = \frac{3}{2} + \frac{\mu_1}{2}
\]
The eigenvalues of the operator $h_2$ is given by the following expression
\begin{equation}\label{eq:h2c124}
 h_2 = q +\frac{3}{2} +\frac{\mu_1}{2} + \mu_2
\end{equation}
The structure function $\Phi^c_1(m)$ is now written
\[
\Phi^c_1(m) = 3 \cdot 2^{15} \hbar^{10} \mu^6  m (m+1) (q+1-m) (m+1 + \mu_1) (m+\mu_1) (q+1-m + 2 \mu_2)
\]
where $m=0,\ldots,q$ and $ \mu_1 > -1, \mu_2 > -1/2.$

Similarly, the structure function for the second \emph{c-system} is given by the following relation
\[
\begin{split}
\Phi^c_2(y)= 3 \cdot 2^9 \hbar^4 \bigl(H_1 - 2 \hbar \mu ( 4 y -1 - \mu_2 ) \bigr) \bigl(H_1 - 2 \hbar \mu ( 4 y -1 + \mu_2 ) \bigr) \\ \bigl(H_1 - 2 \hbar \mu ( 4 y -3 - \mu_2 ) \bigr) \bigl(H_1 - 2 \hbar \mu ( 4 y -3 + \mu_2 ) \bigr) \\ \bigl(H_3 - 4 \hbar \mu ( 2 y-1 - \mu_3 ) \bigr) \bigl(H_3 - 4 \hbar \mu ( 2 y -1 + \mu_3 ) \bigr)
\end{split}
\]
where $ y = x + u.$ Moreover, the eigenvalues of the operator $H_2$  is given using the relation (\ref{eq:Ax})
\[
H_2= \sqrt{\delta} y= 8 \hbar \mu y
\]
Setting
\[
h_1 = \frac{H_1}{2 \hbar \mu}, \quad h_2 = \frac{H_2}{2 \hbar \mu} = 4 y, \quad h_3 = \frac{H_3}{2 \hbar \mu}
\]
the structure function is written
\[
\begin{split}
\Phi^c_2 =  3 \cdot 2^{15} \hbar^{10} \mu^6 \left (  \left(h_2 - h_1 - 1\right)^2 - \mu_2^2   \right ) \left (  \left(h_2 -h_1 - 3\right)^2 - \mu_2^2   \right )\\ \left ( \left(h_3-h_2+2\right)^2 - (2 \mu_3)^2  \right )
\end{split}
\]

From the first \emph{c-system} we have that
\[
h_1 = \frac{3}{2} + \frac{\mu_1}{2} + m, \quad   m=0,\ldots,q
\]
The boundary conditions  $\Phi^c_2(m)=\Phi^c_2(p+1)=0$ where $\Phi^c_2(q)>0$ for all $ q=m+1,\ldots,p$ lead us to the following expressions for the eigenvalues of $h_2$ and $h_3$
\[
h_2 = u + q - m + 3, \quad  u  = q + \frac{3}{2}+ \frac{\mu_1}{2} + \mu_2,  \quad m\leq q \leq p
\]
\begin{equation}\label{eq:h3c124}
h_3 = q + p - m + \frac{7}{2}+ \frac{\mu_1}{2} + \mu_2 + 2 \mu_3
\end{equation}
The function $\Phi^c_2(q)$ is now written
\[
\begin{split}
\Phi^c_2(q) = 3 \cdot 2^{19} \hbar^{10} \mu^6 (q-m)(p+1-q) (q-m+1)(q-m+1 +\mu_2)\\(q-m+\mu_2)(p+1-q+4 \mu_3)
\end{split}
\]
where $\quad m\leq q \leq p$ and $ \mu_2 > -1, \mu_3 > -1/4.$ The  final form  of the eigenvalues for the involving operators composing the generalized anisotropic oscillator chain $(1,2,4)$ is given as follows
\[
H_1 =  \hbar \mu (2 (m+1) + 1  + \mu_1 ), \quad H_2 = \hbar \mu  (4 (q+1) - 2 m + 5 + \mu_1 + 2 \mu_2  )
\]
\[
H_3 = \hbar \mu \bigl(2 (q +  p -  m) + 7 + \mu_1 + 2 \mu_2 + 4 \mu_3 \bigr)
\]
The corresponding structure functions are
\[
\Phi^c_1(m) = 3 \cdot 2^{15} \hbar^{10} \mu^6  m (m+1) (q+1-m) (m+1 + \mu_1) (m+\mu_1) (q+1-m + 2 \mu_2)
\]
\[
\begin{split}
\Phi^c_2(q) = 3 \cdot 2^{19} \hbar^{10} \mu^6 (q-m)(p+1-q) (q-m+1)(q-m+1 +\mu_2)\\(q-m+\mu_2)(p+1-q+4 \mu_3)
\end{split}
\]
\[
m,q,p \in \mathbb{N}, \quad 0\leq m\leq q \leq p,\qquad \mu_1 > -1, \; \mu_2 > -\frac{1}{2}, \mu_3 > - \frac{1}{4}
\]

\section{Four dimensional generalized anisotropic oscillator potential. The $(1,2,4,8)$ system}

The idea of Hamiltonian chains can be applied on four dimensional potentials as well. Let us consider the following sequence of Hamiltonians
\[
H_1 = \frac{1}{2} p_x^2 + k x^2 + \frac{k_1}{x^2}
\]
\[
H_2 = \frac{1}{2} ( p_x^2+ p_y^2 )+ k (x^2 + 4 y^2) + \frac{k_1}{x^2} +  \frac{k_2}{y^2}
\]
\[
H_3 = \frac{1}{2} ( p_x^2+ p_y^2 +p_z^2)+ k (x^2+4 y^2 + 16 z^2) + \frac{k_1}{x^2} +  \frac{k_2}{y^2}+  \frac{k_3}{z^2}
\]
\[
\begin{split}
H_4 = \frac{1}{2} ( p_x^2+ p_y^2 +p_z^2 + p_u^2)+ k (x^2+4 y^2 + 16 z^2+ 64 u^2) + \\ +\frac{k_1}{x^2} +  \frac{k_2}{y^2}+  \frac{k_3}{z^2} + \frac{k_4}{u^2}
\end{split}
\]
The last Hamiltonian is the four dimensional generalized anisotropic oscillator Hamiltonian with coefficients $(a_1,a_2,a_3,a_4)= (1,2,4,8)$

This system implies the existence of three \emph{c-systems}. The first two systems corresponds to the three dimensional $(1,2,4)$ Hamiltonian chain and the third \emph{c-system} is given by the following settings
\begin{equation}\label{eq:subC1248}
x_1 = z,x_2 = u, \kappa = 16 k, \kappa_1 = k_3, \kappa_2 = k_4 , r = H_2
\end{equation}
and we have $\mathcal{H}_1 \equiv H_4 $ and $A_1 \equiv H_3.$
The relation (\ref{eq:generatcubgener}) lead us to the generating function of the corresponding triple cubic Poisson algebra (\ref{eq:cubicclassalgebra}).

The quantum form of the third \emph{c-system} composing the $(1,2,4,8)$ \emph{Hamiltonian chain} can be derived from the expressions (\ref{eq:quantumCsystem}). Furthermore, the form of their quantum algebra as well as the form of the Casimir operator are straightforward calculated by the relations (\ref{eq:AABGenQuanCsyst})-(\ref{eq:CsystCasim}) using the above substitutions (\ref{eq:subC1248}).

For the calculation of the energy eigenvalues we set
\[
k=\frac{\mu^2}{2}, \; k_i = \frac{\hbar^2 (\mu_i^2 - \frac{1}{4})}{2}
\]
the structure function $\Phi^c_3(x)$ for the third \emph{c-system} is written
\[
\begin{split}
\Phi^c_3(x) =  3 \cdot 2^9 \hbar^4 \bigl(H_2 - 4 \hbar \mu  ( 4 z - 1 -\mu_3  ) \bigr )\bigl(H_2 - 4 \hbar \mu  ( 4 z - 1 +\mu_3 ) \bigr ) \\ \bigl(H_2 - 4 \hbar \mu  ( 4 z - 3 -\mu_3 ) \bigr )\bigl(H_2 - 4 \hbar \mu  ( 4 z - 3 +\mu_3 ) \bigr ) \\ \bigl ( H_4 - 8 \hbar \mu (  2 z -1 - \mu_4) \bigr )\bigl ( H_4 - 8 \hbar \mu (  2 z -1 + \mu_4) \bigr )
\end{split}
\]
and eigenvalues of the operator $H_3$ is given by the relation (\ref{eq:ACubRepr})
\[
H_3 =\sqrt{\delta} z= 16 \hbar \mu z
\]
Setting
\[
h_1 = \frac{H_1}{2 \hbar \mu}, \quad h_2 = \frac{H_2}{2 \hbar \mu}, \quad h_3 = \frac{H_3}{2 \hbar \mu} = 8 z, \quad h_4 =\frac{H_4}{2 \hbar \mu}
\]
The structure function $\Phi^c_3 $ is written
\[
\begin{split}
\Phi^c_3 = 3 \cdot 2^{15} \hbar^{10} \mu^6 \left (  \left(h_3 - h_2 - 2\right)^2 - (2 \mu_3)^2   \right ) \left (  \left(h_3 -h_2 - 6 \right)^2 - (2 \mu_3)^2   \right )\\ \left ( \left(h_4-h_3+4\right)^2 - (4 \mu_3)^2  \right )
\end{split}
\]
From the second \emph{c-system} we have that
\[
h_2 = 2 q - m + \frac{9}{2} + \frac{\mu_1}{2} + \mu_2
\]
where $m \leq q \leq p.$

The boundary conditions  $\Phi^c_3(q)=\Phi^c_3(\ell+1)=0$ where $\Phi^c_3(p)>0$ for all $ p=q+1,\ldots,\ell$ lead us to the following formulas for the eigenvalues of the operators $h_3, h_4$
\[
h_3 = w + p +q + 7, \quad  w  = 2 (p - q) - m + \frac{7}{2} + \frac{\mu_1}{2}+ \mu_2 + 2 \mu_3,  \quad q\leq p \leq \ell
\]
\[
h_4 =  2 p - q - m +  \ell + \frac{15}{2}+ \frac{\mu_1}{2} + \mu_2 + 2 \mu_3 + 4 \mu_4
\]
The structure function $\Phi_3^c(p)$ is written
\[
\begin{split}
\Phi_3^c(p) =  9 \cdot 2^{13} \hbar^{10} \mu^6 (p-q)(\ell+1 -p)(3 p - 3 q + 4)(3 p -3 q + 4 + 4 \mu_3) \\(3 p - 3 q + 4 \mu_3) (\ell +1 - p + 8 \mu_4)
\end{split}
\]
where $q \leq p \leq \ell$ and $\mu_3>-7/4, \mu_4 > -1/8$

The  final form  of the eigenvalues for the involving operators composing the generalized anisotropic oscillator chain $(1,2,4,8)$ is given as follows
\[
H_1 =  \hbar \mu (2 (m+1) + 1  + \mu_1 ), \quad H_2 = \hbar \mu  (4 (q+1) - 2 m + 5 + \mu_1 + 2 \mu_2  )
\]
\[
H_3 = \hbar \mu \bigl(2 ( 3 (p+3) - (q+1) -  (m+1)) + 7 + \mu_1 + 2 \mu_2 + 4 \mu_3 \bigr)
\]
\[
H_4 =  \hbar \mu \bigl(2 ( 2 p + \ell -q -  m) + 15 + \mu_1 + 2 \mu_2 + 4 \mu_3 + 8 \mu_4 \bigr)
\]
The corresponding structure functions are
\[
\Phi_1^c(m) = 3 \cdot 2^{15} \hbar^{10} \mu^6  m (m+1) (q+1-m) (m+1 + \mu_1) (m+\mu_1) (q+1-m + 2 \mu_2)
\]
\[
\begin{split}
\Phi_2^c(q) = 3 \cdot 2^{19} \hbar^{10} \mu^6 (q-m)(p+1-q) (q-m+1)(q-m+1 +\mu_2)\\(q-m+\mu_2)(p+1-q+4 \mu_3)
\end{split}
\]
\[
\begin{split}
\Phi_3^c(p) =   9 \cdot 2^{13} \hbar^{10} \mu^6 (p-q)(\ell+1 -p)(3 p - 3 q + 4)(3 p -3 q + 4 + 4 \mu_3) \\(3 p - 3 q + 4 \mu_3) (\ell +1 - p + 8 \mu_4)
\end{split}
\]
where
\[
m,q,p,\ell \in \mathbb{N}, \quad 0\leq m\leq q \leq p \leq \ell ,\quad \mu_1 > -1, \; \mu_2 > -\frac{1}{2}, \mu_3 > - \frac{1}{4}, \mu_4 > - \frac{1}{8}
\]

\section{Discussion}

Using pure algebraic methods of \cite{Das01}, we can calculate the energy eigenvalues of the generalized
anisotropic oscillator chain.
This method can be applied to other Hamiltonian chains in order to produce a more general theory. The study of this kind of algebras, which describe many multidimensional superintegrable systems is an interesting mathematical topic, which is not yet
been explored.

\newpage

\section*{Appendix A. The structure function of the triple cubic algebra for $\beta \neq 0$}

\[
\begin{split}
&\Phi(\mathcal{N}) =  -3 \cdot 2^{16} \beta^8 (\mathcal{N}  + u +1) \bigl( 1 + 2 (\mathcal{N}+u)\bigr) \bigl(1 - 2 (\mathcal{N}+u)\bigr)^2 K + 3 \cdot 2^{13} a^2 \beta^{12} (\mathcal{N}  + u)^{14} - \\&  - 3^3 \cdot 2^{12} a^2 \beta^{12} (\mathcal{N}  + u)^{13} + 2^{14} \beta^{10} \bigl ( a (6 \alpha \beta + 7 a \beta^2 - 9a \delta) + 3 \mu \beta^2 \bigr )(\mathcal{N}  + u)^{12} + \\&   + 3 \cdot 2^{11} \beta^{10} \bigl ( a (47 a \beta^2 - 56 \alpha \beta + 84 a \delta) -28 \mu \beta^2  \bigr ) (\mathcal{N}  + u)^{11} -512 \beta ^8 \Bigl (-192 \alpha ^2 \beta ^2 + \\&  +a^2 (775 \beta ^4+648 \beta ^2 \delta -720 \delta ^2)+48 \alpha  a \beta
    (20 \delta -9 \beta ^2)-384 a \beta ^2 \gamma - \\&  -8 \beta ^2 (27 \beta ^2 \mu +32 \beta  \nu -48 \delta  \mu
   ) \Bigr ) (\mathcal{N}  + u)^{10} -256 \beta ^8 \Bigl ( 960 \alpha ^2 \beta ^2+a^2 (1261 \beta ^4+4440 \beta ^2 \delta + \\ &  +3600 \delta ^2)-80 \alpha  a
   \beta  (37 \beta ^2+60 \delta )+1920 a \beta ^2 \gamma -40 \beta ^2 (37 \beta ^2 \mu -32 \beta  \nu +48
   \delta  \mu )\Bigr) (\mathcal{N}  + u)^{9} + \\&  +256 \beta ^6 \Bigl(192 \alpha ^2 (\beta ^4-8 \beta ^2 \delta )+a^2 (1849 \beta ^6+3888 \beta ^4 \delta +720
   \beta ^2 \delta ^2-1920 \delta ^3)- \\&  -96 \alpha  (a (27 \beta ^5+10 \beta ^3 \delta -40 \beta  \delta
   ^2)-  16 \beta ^3 \gamma )+384 a \beta ^2 (\beta ^2 \gamma +4 \beta  \epsilon -8 \gamma  \delta )- \\ & - 16
   \beta ^2 (81 \beta ^4 \mu -16 \beta ^3 \nu +24 \beta ^2 (\delta  \mu -4 \xi )+96 \beta  \delta  \nu -72 \delta ^2 \mu
   )\Bigr) (\mathcal{N}  + u)^{8} + \\ & + 2304 \beta ^6 \Bigl(64 \alpha ^2 (3 \beta ^4+4 \beta ^2 \delta )+a^2 (91 \beta ^6+396 \beta ^4 \delta +720
   \beta ^2 \delta ^2+320 \delta ^3)-8 \alpha  (a (33 \beta ^5 + \\ & +120 \beta ^3 \delta +80 \beta  \delta
   ^2)+32 \beta ^3 \gamma )+128 a \beta ^2 (3 \beta ^2 \gamma -2 \beta  \epsilon +4 \gamma  \delta )-4
   \beta ^2 (33 \beta ^4 \mu -64 \beta ^3 \nu + \\ & +32 \beta ^2 (3 \delta  \mu +2 \xi )-64 \beta  \delta  \nu +48 \delta ^2 \mu
   )\Bigr) (\mathcal{N}  + u)^{7}  -32 \beta ^4 \Bigl(128 \alpha ^2 (51 \beta ^6-56 \beta ^4 \delta -144 \beta ^2 \delta ^2)+ \\ & + a^2 (8779 \beta
   ^8+24720 \beta ^6 \delta +24480 \beta ^4 \delta ^2-8960 \beta ^2 \delta ^3-11520 \delta ^4)-32 \alpha  (5 a
   (103 \beta ^7+204 \beta ^5 \delta - \\ & -112 \beta ^3 \delta ^2-192 \beta  \delta ^3)-32 \beta ^3 (7 \beta ^2
   \gamma -24 \beta  \epsilon +36 \gamma  \delta ))+256 a \beta ^2 (51 \beta ^4 \gamma +28 \beta ^3 \epsilon
   -56 \beta ^2 \gamma  \delta + \\ &  + 144 \beta  \delta  \epsilon  -144 \gamma  \delta ^2)-16 \beta ^2 (515 \beta ^6 \mu
   -544 \beta ^5 \nu +16 \beta ^4 (51 \delta  \mu -28 \xi )+64 \beta ^3 (7 \delta  \nu +48 \zeta )+ \\ & + 48 \beta ^2 (16 \gamma
   ^2-7 \delta ^2 \mu -32 \delta  \xi )+768 \beta  \delta ^2 \nu -384 \delta ^3 \mu )\Bigr) (\mathcal{N}  + u)^{6} - \\ & - 48 \beta ^4 \Bigl(128 \alpha ^2 (39 \beta ^6+136 \beta ^4 \delta +48 \beta ^2 \delta ^2)+a^2 (1559 \beta
   ^8+7056 \beta ^6 \delta +18720 \beta ^4 \delta ^2 + \\ & +21760 \beta ^2 \delta ^3+3840 \delta ^4)-32 \alpha  (a
   (147 \beta ^7+780 \beta ^5 \delta +1360 \beta ^3 \delta ^2+320 \beta  \delta ^3)+ \\ & +32 \beta ^3 (17 \beta ^2
   \gamma -8 \beta  \epsilon +12 \gamma  \delta ))+256 a \beta ^2 (39 \beta ^4 \gamma -68 \beta ^3 \epsilon
   +136 \beta ^2 \gamma  \delta -48 \beta  \delta  \epsilon +48 \gamma  \delta ^2)-\\ & -16 \beta ^2 (147 \beta ^6 \mu
   -416 \beta ^5 \nu +16 \beta ^4 (39 \delta  \mu +68 \xi )-64 \beta ^3 (17 \delta  \nu +16 \zeta )+ \\ & + 16 \beta ^2 (-16
   \gamma ^2+51 \delta ^2 \mu +32 \delta  \xi )-256 \beta  \delta ^2 \nu +128 \delta ^3 \mu )\Bigr) (\mathcal{N}  + u)^{5} + \\ &  +  32 \beta ^2 \Bigl(768 \alpha ^2 (5 \beta ^8+7 \beta ^6 \delta -24 \beta ^4 \delta ^2-16 \beta ^2 \delta ^3)+a^2
   (2749 \beta ^{10}+8766 \beta ^8 \delta +12352 \beta ^6 \delta ^2 + \\& +4672 \beta ^4 \delta ^3-11520 \beta ^2 \delta ^4-4608
   \delta ^5)-4 \alpha  (a (1461 \beta ^9+4800 \beta ^7 \delta +2336 \beta ^5 \delta ^2-7680 \beta ^3 \delta
   ^3 -\\ & -3840 \beta  \delta ^4)+192 \beta ^3 (32 \beta  \epsilon  (\beta ^2+2 \delta )+\gamma  (7 \beta
   ^4-48 \beta ^2 \delta -48 \delta ^2)))+256 a \beta ^2 (30 \beta ^6 \gamma -53 \beta ^5 \epsilon + \\ & + 42
   \beta ^4 \gamma  \delta +144 \beta ^3 \delta  \epsilon -144 \beta ^2 \gamma  \delta ^2+144 \beta  \delta ^2 \epsilon -96
   \gamma  \delta ^3)-2 \beta ^2 (1461 \beta ^8 \mu -2560 \beta ^7 \nu + \\ & +384 \beta ^6 (10 \delta  \mu +7 \xi )+384
   \beta ^5 (64 \zeta -7 \delta  \nu )+32 \beta ^4 (320 \gamma ^2+\delta  (31 \delta  \mu -256 \xi ))- \\ & -6144 \beta ^3
   (4 \gamma  \epsilon -\delta  (\delta  \nu +4 \zeta ))-3072 \beta ^2 \delta  (\delta  (\delta  \mu +2 \xi )-4 \gamma
   ^2)+2048 \beta  \delta ^3 \nu -768 \delta ^4 \mu )\Bigr) (\mathcal{N}  + u)^{4} +
\end{split}
\]
\[
\begin{split}
& + 8 \beta ^2 \Bigl(384 \alpha ^2 \beta ^2 (\beta ^2+4 \delta ) (17 \beta ^4+32 \beta ^2 \delta -16 \delta
   ^2)+a^2 (1661 \beta ^{10}+7404 \beta ^8 \delta +20384 \beta ^6 \delta ^2+43904 \beta ^4 \delta ^3 + \\ &  +26880 \beta ^2
   \delta ^4-9216 \delta ^5)-8 \alpha  (a (617 \beta ^9+4080 \beta ^7 \delta +10976 \beta ^5 \delta ^2+8960
   \beta ^3 \delta ^3-3840 \beta  \delta ^4)+ \\ & + 64 \beta ^3 (75 \beta ^4 \gamma -112 \beta ^3 \epsilon +168 \beta ^2
   \gamma  \delta +192 \beta  \delta  \epsilon -144 \gamma  \delta ^2))+256 a \beta ^2 (51 \beta ^6 \gamma
   -214 \beta ^5 \epsilon +300 \beta ^4 \gamma  \delta - \\ & -336 \beta ^3 \delta  \epsilon +336 \beta ^2 \gamma  \delta ^2+ 288 \beta
    \delta ^2 \epsilon -192 \gamma  \delta ^3)+4 \beta ^2 (-617 \beta ^8 \mu +2176 \beta ^7 \nu -192 \beta ^6 (17
   \delta  \mu +50 \xi )+ \\ & + 128 \beta ^5 (75 \delta  \nu +224 \zeta )+32 \beta ^4 (96 \gamma ^2-193 \delta ^2 \mu -576 \delta
    \xi )+1024 \beta ^3 (24 \gamma  \epsilon +7 \delta ^2 \nu -24 \delta  \zeta )-\\&-512 \beta ^2 \delta
   (24 \gamma ^2+7 \delta ^2 \mu -12 \delta  \xi )-2048 \beta  \delta ^3 \nu +768 \delta ^4 \mu )\Bigr) (\mathcal{N}  + u)^{3} - \\ & -2 \Bigl(192 \alpha ^2 \beta ^2 (\beta ^2+4 \delta ) (71 \beta ^6-12 \beta ^4 \delta -176 \beta ^2 \delta ^2-64
   \delta ^3)+a^2 (6929 \beta ^{12}+23400 \beta ^{10} \delta +26544 \beta ^8 \delta ^2 + \\ & +40704 \beta ^6 \delta
   ^3-26880 \beta ^4 \delta ^4-55296 \beta ^2 \delta ^5-12288 \delta ^6)-16 \alpha  (a (975 \beta ^{11}+4260
   \beta ^9 \delta +5088 \beta ^7 \delta ^2- \\ & -4480 \beta ^5 \delta ^3-11520 \beta ^3 \delta ^4-3072 \beta  \delta ^5)+64
   \beta ^3 (3 \gamma  (\beta ^2-4 \delta ) (17 \beta ^4+40 \beta ^2 \delta +16 \delta ^2)+\\ & + 8 \beta
   \epsilon  (7 \beta ^4+72 \beta ^2 \delta +48 \delta ^2)))+384 a \beta ^2 (\gamma  (\beta
   ^2+4 \delta ) (71 \beta ^6-12 \beta ^4 \delta -176 \beta ^2 \delta ^2-64 \delta ^3) + \\ & + 8 \beta  \epsilon
   (-49 \beta ^6+28 \beta ^4 \delta +144 \beta ^2 \delta ^2+64 \delta ^3))-8 \beta ^4 (975 \beta ^8 \mu
   -2272 \beta ^7 \nu +48 \beta ^6 (71 \delta  \mu +136 \xi )+\\ & + 128 \beta ^5 (112 \zeta -51 \delta  \nu )+32 \beta ^4 (496
   \gamma ^2+\delta  (57 \delta  \mu +160 \xi ))-512 \beta ^3 (144 \gamma  \epsilon -\delta  (7 \delta  \nu +144 \zeta
   ))+ \\ & + 256 \beta ^2 (144 \gamma ^2 \delta -7 \delta ^3 \mu -72 \delta ^2 \xi +384 \epsilon ^2)+6144 \beta  \delta
   (\delta ^2 \nu -16 \gamma  \epsilon )+768 (32 \gamma ^2 \delta ^2-3 \delta ^4 \mu ))\Bigr)  (\mathcal{N}  + u)^{2} - \\ & -192 \alpha ^2 \Bigl(\beta ^3+4 \beta  \delta )^2 (21 \beta ^4+8 \beta ^2 \delta -48 \delta ^2)-a^2
   (\beta ^2+4 \delta ) (915 \beta ^{10}+300 \beta ^8 \delta +5728 \beta ^6 \delta ^2+ \\ & +11136 \beta ^4 \delta
   ^3-2304 \beta ^2 \delta ^4-9216 \delta ^5)+16 \alpha  \beta  ((\beta ^2+4 \delta ) (a (165
   \beta ^8+600 \beta ^6 \delta +1856 \beta ^4 \delta ^2-384 \beta ^2 \delta ^3-\\&-2304 \delta ^4)+192 \beta ^2 \gamma
   (11 \beta ^4-48 \delta ^2))-512 \beta ^3 \epsilon  (11 \beta ^4-24 \beta ^2 \delta -144 \delta
   ^2))+128 a \beta ^2 (8 \beta  \epsilon  (65 \beta ^6+\\ & + 132 \beta ^4 \delta -144 \beta ^2 \delta ^2-576
   \delta ^3)-3 \gamma (\beta ^2+4 \delta )^2 (21 \beta ^4+8 \beta ^2 \delta -48 \delta
   ^2))+8 \beta ^4 (165 \beta ^8 \mu -672 \beta ^7 \nu +\\ & + 48 \beta ^6 (21 \delta  \mu +88 \xi )-1408 \beta ^5 (3
   \delta  \nu +16 \zeta )-32 \beta ^4 (48 \gamma ^2-\delta  (67 \delta  \mu +480 \xi ))-\\& - 512 \beta ^3 (48 \gamma
   \epsilon +\delta  (11 \delta  \nu -48 \zeta ))+256 \beta ^2 (48 \gamma ^2 \delta +11 \delta ^3 \mu -24 \delta ^2 \xi
   +1152 \epsilon ^2)+2048 \beta  \delta  (\delta ^2 \nu -144 \gamma  \epsilon )+\\ & + 768 (96 \gamma ^2 \delta
   ^2-\delta ^4 \mu )\Bigr)  (\mathcal{N}  + u) + \\ & + 64 \alpha ^2 \Bigl(\beta ^3+4 \beta  \delta )^2 (33 \beta ^4-56 \beta ^2 \delta +16 \delta ^2)+a^2 (\beta
   ^2+4 \delta ) (865 \beta ^{10}-460 \beta ^8 \delta +1568 \beta ^6 \delta ^2+2176 \beta ^4 \delta ^3-\\& -4864 \beta ^2
   \delta ^4+1024 \delta ^5)-16 \alpha  \beta  (\beta ^2+4 \delta ) (a (125 \beta ^8+160 \beta ^6
   \delta +416 \beta ^4 \delta ^2-1024 \beta ^2 \delta ^3+256 \delta ^4)+\\& + 64 \beta ^2 \gamma  (13 \beta ^4-40 \beta
   ^2 \delta +16 \delta ^2))+128 a \beta ^2 (3 \beta ^2-4 \delta ) (\gamma  (11 \beta ^2-4
   \delta ) (\beta ^2+4 \delta )^2-\\ &-8 \beta  \epsilon  (15 \beta ^4+24 \beta ^2 \delta -16 \delta
   ^2))+8192 \alpha  \beta ^4 \epsilon  (\beta ^4-24 \beta ^2 \delta +16 \delta ^2)-8 \beta ^4 (125
   \beta ^8 \mu -352 \beta ^7 \nu +\\ & + 16 \beta ^6 (33 \delta  \mu +104 \xi )-128 \beta ^5 (13 \delta  \nu +16 \zeta )+32 \beta ^4
   (112 \gamma ^2+\delta  (7 \delta  \mu +160 \xi ))-\\ & - 512 \beta ^3 (48 \gamma  \epsilon +\delta  (\delta  \nu -48
   \zeta ))+256 \beta ^2 (48 \gamma ^2 \delta +\delta ^3 \mu -24 \delta ^2 \xi -128 \epsilon ^2)+ \\ & +2048 \beta
   (16 \gamma  \delta  \epsilon +\delta ^3 \nu )-256 (32 \gamma ^2 \delta ^2+3 \delta ^4 \mu )\Bigr)
\end{split}
\]

\end{document}